\documentclass[aps,prd,superscriptaddress,showpacs,showkeys,floatfix,twocolumn,nofootinbib]{revtex4}

\usepackage{amsmath,amssymb,amsfonts,dcolumn,slashed,dsfont}
\usepackage{graphicx}
\usepackage{psfrag}

\newcommand{\mpi}{M_{\pi}}

\newcommand{\meta}{M_\eta}
\newcommand{\Fpi}{F_\pi}
\newcommand{\gA}{g_A}
\newcommand{\beq}{\begin{equation}}
\newcommand{\eeq}{\end{equation}}
\newcommand{\mc}{m_\chi}
\newcommand{\mN}{m_N}
\newcommand{\mA}{m_A}
\newcommand{\muN}{\mu_N}
\newcommand{\muA}{\mu_A}
\newcommand{\mpp}{m_p}
\newcommand{\mn}{m_n}
\newcommand{\muu}{m_u}
\newcommand{\md}{m_d}
\newcommand{\qq}{\mathbf{q}}
\newcommand{\pp}{\mathbf{p}}
\newcommand{\PP}{\mathbf{P}}
\newcommand{\KK}{\mathbf{K}}
\newcommand{\vv}{\mathbf{v}}
\newcommand{\vvp}{\mathbf{v}^\perp}
\newcommand{\N}{\mathcal{N}}
\newcommand{\M}{\mathcal{M}}
\newcommand{\F}{\mathcal{F}}
\newcommand{\spin}{\mathbf{S}}
\newcommand{\unity}{\mathds{1}}
\newcommand{\sig}{\boldsymbol{\sigma}}
\newcommand{\ttau}{\boldsymbol{\tau}}
\newcommand{\diff}{\text{d}}

\newcommand{\Order}{\mathcal{O}}
\newcommand{\Op}{\mathcal{O}}

\providecommand{\MeV}{\,\text{MeV}}
\providecommand{\GeV}{\,\text{GeV}}
\providecommand{\fm}{\,\text{fm}}

\allowdisplaybreaks[1]

\begin{document}


\title{Analysis strategies for general spin-independent WIMP--nucleus scattering}

\author{Martin Hoferichter}
\email[E-mail:~]{mhofer@uw.edu}
\affiliation{Institute for Nuclear Theory, University of Washington, Seattle, WA 98195-1550, USA}

\author{Philipp Klos}
\email[E-mail:~]{pklos@theorie.ikp.physik.tu-darmstadt.de}
\affiliation{Institut f\"ur Kernphysik, 
Technische Universit\"at Darmstadt, 
64289 Darmstadt, Germany}
\affiliation{ExtreMe Matter Institute EMMI, 
GSI Helmholtzzentrum f\"ur Schwerionenforschung GmbH, 
64291 Darmstadt, Germany}

\author{Javier Men\'endez}
\email[E-mail:~]{menendez@nt.phys.s.u-tokyo.ac.jp}
\affiliation{Department of Physics, The University of Tokyo, 113-0033 Tokyo, Japan}

\author{Achim Schwenk}
\email[E-mail:~]{schwenk@physik.tu-darmstadt.de}
\affiliation{Institut f\"ur Kernphysik, 
Technische Universit\"at Darmstadt, 
64289 Darmstadt, Germany}
\affiliation{ExtreMe Matter Institute EMMI, 
GSI Helmholtzzentrum f\"ur Schwerionenforschung GmbH, 
64291 Darmstadt, Germany}
\affiliation{Max-Planck-Institut f\"ur Kernphysik, Saupfercheckweg 1, 
69117 Heidelberg, Germany}

\begin{abstract}

We propose a formalism for the analysis of direct-detection
dark-matter searches that covers all coherent responses for scalar and
vector interactions and incorporates QCD constraints imposed by chiral
symmetry, including all one- and two-body WIMP--nucleon interactions
up to third order in chiral effective field theory.  One of the free
parameters in the WIMP--nucleus cross section corresponds to standard
spin-independent searches, but in general different combinations of
new-physics couplings are probed.  We identify the interference with
the isovector counterpart of the standard spin-independent response
and two-body currents as the dominant corrections to the leading
spin-independent structure factor, and discuss the general
consequences for the interpretation of direct-detection experiments,
including minimal extensions of the standard spin-independent
analysis.  Fits for all structure factors required for the scattering
off xenon targets are provided based on state-of-the-art nuclear
shell-model calculations.

\end{abstract}



\maketitle

\section{Introduction}
\label{sec:intro}

Direct searches for the nuclear recoil produced by weakly interacting
massive particles (WIMPs) on target nuclei in large-scale detectors
provide a prime avenue to unravel the nature of dark matter,
complementary to indirect searches for annihilation remnants in
astrophysical observations and the production of dark-matter particles
in collider experiments~\cite{Baudis:2016qwx}.  However, for the
interpretation of current experimental limits,
e.g.,~\cite{Aprile:2012nq,Agnese:2014aze,Agnese:2015nto,Akerib:2015rjg,%
Xiao:2015psa,Angloher:2015ewa,Amole:2016pye,Agnes:2015ftt,Armengaud:2016cvl}
it is crucial that the nuclear aspects of direct-detection experiments
be adequately addressed.  This is especially important given the
impressive experimental efforts that include future liquid-noble-gas
ton-scale experiments already in commissioning such as
XENON1T~\cite{Aprile:2015uzo}, DEAP-3600~\cite{Boulay:2012hq}, and
ArDM~\cite{Calvo:2015uln}, or in planning phase,
LZ~\cite{Akerib:2015cja}, XENONnT~\cite{Aprile:2014zvw},
XMASS~\cite{Liu:2014hda}, DarkSide-20k~\cite{DarkSideproject:2016zfx},
and DARWIN~\cite{Aalbers:2016jon}; but also smaller-scale experiments
such as SuperCDMS SNOLAB~\cite{Brink:2012zza},
DAMIC100~\cite{Chavarria:2014ika}, or CRESST~\cite{Angloher:2015eza}
that focus on light WIMPs with masses below $10\GeV$.

Standard analyses of WIMP--nucleus scattering are formulated in terms
of spin-independent (SI) and spin-dependent (SD)
searches~\cite{Engel:1992bf}, named after the nature of the
WIMP--nucleon interactions at low energies.  At the same time, SI and
SD scattering are characterized by a very different scaling of the
corresponding structure factors: while for SI scattering the response
is proportional to the total number of nucleons $A^2$, the scale of SD
scattering is set by the spin expectation value of the unpaired
nucleon.  Due to the coherent enhancement of SI interactions, the
corresponding limits on the WIMP--nucleon couplings set by
direct-detection experiments are orders of magnitude more stringent
than for SD searches, but each type of interaction is sensitive to
different operators for the coupling of WIMPs with Standard-Model
fields.  For instance, while quark--WIMP scalar--scalar and
vector--vector terms contribute to the SI response, the SD interaction
is generated by axial-vector--axial-vector operators.  Additional
information on the WIMP nature can be extracted from inelastic
scattering off the target nuclei~\cite{Baudis:2013bba,McCabe:2015eia}.

Corrections to standard SI and SD responses are conveniently studied
in terms of effective field theories (EFTs). In this context, the
calculation of nuclear structure factors has been organized in two
different ways: first, non-relativistic EFT (NREFT) for nucleon and
WIMP
fields~\cite{Fan:2010gt,Fitzpatrick:2012ix,Fitzpatrick:2012ib,Anand:2013yka}
allows a study of the nuclear responses as a function of the effective
couplings in the EFT, and to extract limits on the coefficients of the
NREFT operators~\cite{Schneck:2015eqa}.  Second, in order to translate
the NREFT limits to the parameter space of a given new-physics model,
the QCD dynamics integrated out in the NREFT approach needs to be
included.  Particularly important are the consequences of the
spontaneous breaking of the chiral symmetry of QCD, which can be
explored within the framework of chiral EFT (ChEFT), see
Refs.~\cite{Epelbaum:2008ga,Machleidt:2011zz,Hammer:2012id,Bacca:2014tla}
for recent reviews.  The analysis within ChEFT establishes relations
between different NREFT operators, and provides a counting scheme that
indicates at which order contributions beyond the single-nucleon
level~\cite{Prezeau:2003sv,Cirigliano:2012pq,Menendez:2012tm,Hoferichter:2015ipa}
need to be included. Recent work in this direction includes
ChEFT-based structure factors for the SD
response~\cite{Menendez:2012tm,Klos:2013rwa}, aspects of SI
scattering~\cite{Cirigliano:2012pq,Cirigliano:2013zta,Vietze:2014vsa},
inelastic scattering~\cite{Baudis:2013bba}, as well as a general ChEFT
analysis of one- and two-body currents~\cite{Hoferichter:2015ipa}.

In the present work we provide a generalization of SI scattering that
includes all coherent contributions up to third order in
ChEFT~\cite{Hoferichter:2015ipa}.  This involves considering two-body
currents, but also momentum corrections to the nucleon form factors
predicted at the same ChEFT order. 
We provide a detailed discussion
of the structure factor associated with the scalar two-body current
studied before in the literature, and extend the analysis
to include the two-body current generated by the coupling of the trace anomaly of
the QCD energy-momentum tensor to the pion in flight, which becomes 
important if the WIMP couples (significantly) via gluonic interactions.
In addition, an analysis of the
NREFT operators reveals that in general there are six relevant nuclear
operators, denoted by $M$, $\Sigma'$, $\Sigma''$, $\Delta$,
$\tilde\Phi'$, and
$\Phi''$~\cite{Fitzpatrick:2012ix,Anand:2013yka},
where $M$ corresponds to the standard SI scattering, while a
combination of $\Sigma'$ and $\Sigma''$ yields the operator relevant
for the SD case.  Given that apart from $M$ also $\Phi''$ can be
coherently enhanced (which is especially the case in heavy nuclei) and
that $M$ and $\Phi''$ interfere, a generalization of the traditional
SI analysis should also take the effects from $\Phi''$ into
account~\cite{Fitzpatrick:2012ix}.

We note that for general SI scattering, new combinations of Wilson
coefficients are probed by the two-body currents coupling to the
exchanged pion in flight, and also by the corrections to the nucleon
form factors and the contributions associated with the $\Phi''$
operator. This is in contrast to the SD case, where the dominant
two-body currents can be absorbed into a redefinition of the one-body
structure factors, i.e., the two-body correction is sensitive to the
same physics beyond the Standard Model (BSM) as the standard SD
interaction~\cite{Menendez:2012tm,Klos:2013rwa}. 
In a similar way to the SI analysis presented here, a more general SD
analysis should include the effects of all relevant nuclear operators and two-body currents.

This work is organized as follows. We start with an overview of the
main results in Sec.~\ref{sec:overview}, where we propose an analysis
strategy for direct-detection experiments that generalizes the
standard SI case. The general formalism is detailed in
Sec.~\ref{sec:formalism}, where we lay out the decomposition of the
WIMP--nucleus scattering rate, collect the relevant nucleon matrix
elements, and introduce the Wilson coefficients that parameterize the
WIMP--quark and WIMP--gluon interactions. We then formulate a set of
generalized structure factors that includes effects from two-body
currents, corrections to the nucleon form factors, and the nuclear
$\Phi''$ operator.  In Sec.~\ref{sec:one-body} we present
state-of-the-art nuclear shell-model calculations for the structure
factors corresponding to one-body currents in all relevant xenon
isotopes, before developing a generalization for the two-body currents
in Sec.~\ref{sec:two-body}.  In Sec.~\ref{sec:parameters_radius} we
discuss the size of the nucleon form-factor corrections as well as the
number of independent parameters in generalized SI scattering,
and work out in detail the size of the corrections to standard SI scattering
for two simple models. We conclude with a short summary in Sec.~\ref{sec:summary}. 
While our analysis strategy is general, the numerical results presented
here are focused on WIMPs scattering off xenon nuclei,
leaving the nuclear structure calculations for other targets to future work.

\section{Overview of main results and analysis strategies}
\label{sec:overview}

Standard analyses of dark-matter direct-detection experiments
distinguish between SI and SD scattering based on the nature of the
WIMP--nucleon interaction.  At the same time, these two cases generate
very different nuclear responses, as SI scattering is enhanced by the
coherent contribution of all nucleons in the nucleus, whereas the scale
of SD scattering is set by a single-nucleon matrix element.

When subleading contributions in EFTs are considered, the
classification of the different terms according to the nature of the
WIMP--nucleon interaction becomes less useful, given that the coherent
enhancement associated with the combined contribution of a significant
number of nucleons is also common to NREFT operators that may involve
a WIMP or even a nucleon spin operator.  Such responses are closer in
their experimental signature to the traditional SI interactions in the
sense that the associated structure factors are enhanced compared to
the single-nucleon case.

Therefore we propose to define generalized SI scattering not by the
form of the NREFT operator, but based on whether a coherent
enhancement is possible.  In this spirit, a general decomposition of
the WIMP--nucleus cross section $\sigma_{\chi\N}^\text{SI}$ should
include the coherently-enhanced corrections generated by
\begin{enumerate}
 \item the standard SI isoscalar WIMP--nucleon interaction,
 \item its isovector counterpart,
 \item the interaction of the WIMP with two nucleons via two-body (meson-exchange) currents,
 \item momentum-dependent corrections to the nucleon form factors,
 \item the quasi-coherent response associated with the $\Phi''$ operator
 (related to the nucleon spin-orbit operator).
\end{enumerate}

\begin{figure*}[t] 
\centering
\includegraphics[width=.9\textwidth,clip]{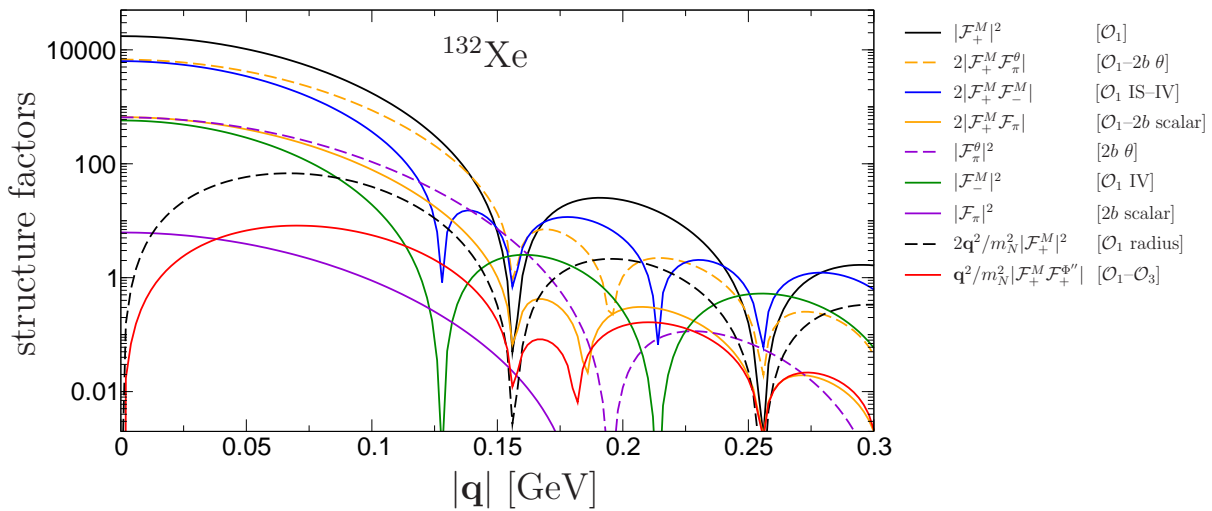}
\caption{Comparison of the leading structure factors associated with
the coherent and quasi-coherent one-body $\F^M$ and $\F^{\Phi''}$
nuclear responses, the two-body nuclear responses $\F_\pi$ (solid lines) and $\F_\pi^\theta$ (dashed lines), and the
radius corrections ($\dot c$) to the structure factors.  The
individual contributions are ordered in the legend according to their
size at $|\qq|=0$ (from top to bottom):
the standard SI response $\F^M_+$ corresponding to
the isoscalar one-body $\Op_1$ operator (black), its interference with
an $\Op_1$ isovector contribution (blue) and with the two-body
responses $\F_\pi$ and $\F_\pi^\theta$ (orange), the purely isovector contribution $\F_-^M$
(green) and the structure factor generated solely by the two-body
currents (violet), the momentum-dependent radius correction
to $\Op_1$ (black dashed), and the interference of the standard SI
response with the quasi-coherent one-body $\F^{\Phi''}$ structure factor (red). 
The results, representative for all stable xenon isotopes, are shown for
the most abundant $^{132}$Xe.}
\label{fig:sf_overview}
\end{figure*}

The proposed generalization amounts to the decomposition of the
WIMP--nucleus cross section
\begin{align}
\label{cross_section_dec}
\frac{\diff \sigma_{\chi\N}^\text{SI}}{\diff \qq^2}&=\frac{1}{4\pi\vv^2}\bigg|\Big(c_+^M-\frac{\qq^2}{m_N^2} \, \dot c_+^M\Big)\F_+^M(\qq^2)+c_\pi \F_\pi(\qq^2)\notag\\
&+c_\pi^\theta \F_\pi^\theta(\qq^2)+\Big(c_-^M-\frac{\qq^2}{m_N^2} \, \dot c_-^M\Big)\F_-^M(\qq^2)\notag\\
&+\frac{\qq^2}{2\mN^2}\Big[c_+^{\Phi''}\F_+^{\Phi''}(\qq^2)+c_-^{\Phi''}\F_-^{\Phi''}(\qq^2)\Big]\bigg|^2,
\end{align}
where $\qq$ is the momentum transfer, $\vv$ the WIMP velocity, and, generically,
the nuclear responses are denoted by $\F$ and the free parameters that
include BSM physics by $c$.  This cross section includes all coherent
contributions mentioned above and all terms up to third order in
ChEFT~\cite{Hoferichter:2015ipa}.  First, the standard SI nuclear
$\F^M$ response, associated with the NREFT operator $\Op_1$
[see Eq.~\eqref{Op_Wick} for definitions of the NREFT $\Op_i$ operators],
can be sensitive to protons and neutrons in the same way (isoscalar, $+$), as
considered in standard SI analyses, but also in the opposite way
(isovector, $-$). Given that the heavy nuclei typically used for
direct-detection experiments have a substantial neutron excess, the
resulting isovector structure factor is coherently enhanced as well.
Next, the power counting of ChEFT predicts to this order two-body
interactions (parameterized by the nuclear $\F_\pi$ and $\F_\pi^\theta$ responses for
the coupling to the pion via a scalar current and via the trace anomaly of the QCD energy-momentum tensor $\theta^\mu_\mu$, respectively) 
and momentum-dependent corrections to
$\Op_1$ (represented by $\dot c$), both of which are coherent.
Finally, contributions from subleading NREFT operators can also be
significantly coherent, the most relevant being $\Op_3$, which is
related to the nucleon spin-orbit operator and gives rise to the
nuclear $\F^{\Phi''}$ response.  Here the coherence is also found in
both isoscalar and isovector cases.

Equation~\eqref{cross_section_dec} reflects the different particle,
hadronic, and nuclear scales involved in WIMP--nucleus scattering.
Within a given new-physics model, WIMPs interact with quark and gluon
degrees of freedom, which are then to be embedded into the nucleon sector.
In an EFT approach the BSM interaction is encoded in the Wilson
coefficients of effective operators, while the nucleon matrix elements
are decomposed into nucleon form factors.  As a result, the free
coefficients $c_\pm^M$, $c_\pi$, $c_\pi^\theta$, $\dot c_\pm^M$, and $c_\pm^{\Phi''}$
correspond to a convolution of Wilson coefficients and nucleon matrix
elements.  In a final step, the nuclear responses $\F_\pm^M$,
$\F_\pi$, $\F_\pi^\theta$, and $\F_\pm^{\Phi''}$ take into account that the scattering
occurs in the nucleus, a strongly interacting many-nucleon system.  In
this work, the relation between the free parameters $c_\pm^M$,
$c_\pi$, $c_\pi^\theta$, $\dot c_\pm^M$, $c_\pm^{\Phi''}$ and the BSM Wilson
coefficients is worked out in Sec.~\ref{sec:formalism} for the case of
a spin-$1/2$ WIMP, see also Eqs.~\eqref{cMpm}--\eqref{cPhipm} for the
explicit relations.  The nuclear responses $\F_\pm^M$, $\F_\pi$, $\F_\pi^\theta$, and
$\F_\pm^{\Phi''}$ are calculated in the framework of the nuclear shell
model, with fit functions given for all stable xenon isotopes in
Sec.~\ref{sec:one-body} for one-body currents and in
Sec.~\ref{sec:two-body} for two-body currents.

The size of the individual terms in Eq.~\eqref{cross_section_dec}
depends on a given new-physics model, which, together with the
nucleon matrix elements, fixes the coefficients $c$.  Nevertheless
the nuclear responses $\F$ already imply a strong hierarchy by
themselves. This is illustrated in Fig.~\ref{fig:sf_overview}, where
the different structure factors including interference terms are
compared under the assumption that all coefficients are the same.  As
expected, the dominant correction originates from the interference of
isoscalar and isovector $\F^M_{\pm}$ responses. Next in the
hierarchy is the interference with the two-body responses $\F_\pi$ and $\F_\pi^\theta$.  The
additional corrections included in Fig.~\ref{fig:sf_overview} (apart
from the pure isovector $\F_-$ and pure two-body contributions) vanish
at $|\qq|=0$, and are therefore suppressed at small $\qq$ compared to
$\Op_1$ and the two-body structure factors. We have also considered further
higher-order NREFT one-body operators, but their contribution is even
more suppressed, see Secs.~\ref{sec:structure} and~\ref{sec:one-body}.
Let us emphasize again that the hierarchy of the structure factors in Fig.~\ref{fig:sf_overview} assumes
a common value for the $c$ coefficients, but these are not in general
independent and relative suppressions or enhancements may occur. In Sec.~\ref{sec:scalar_ud} 
we study the relative size of the
isovector and two-body contributions in two simple models, which for instance
suggests that the large $\F_\pi^\theta$ structure factor tends to be
compensated by a large single-nucleon matrix element, leading to a relative
two-body effect similar to that of the $\F_\pi$ contribution.

Despite the potential impact of the $c$ coefficients on the measured rate, the hierarchy of the nuclear structure factors observed
in Fig.~\ref{fig:sf_overview} is sufficiently pronounced to motivate a minimal extension of the
standard SI scattering of the form
\begin{align}
\label{cross_section_dec_min}
\frac{\diff \sigma_{\chi\N}^\text{SI}}{\diff \qq^2}&=\frac{1}{4\pi\vv^2}\Big|c_+^M\F_+^M(\qq^2)+c_-^M\F_-^M(\qq^2)\notag\\
&+c_\pi \F_\pi(\qq^2)+c_\pi^\theta \F_\pi^\theta(\qq^2)\Big|^2,
\end{align}
with only $4$ independent parameters.

Since the nuclear responses can be obtained from nuclear-structure
calculations, direct-detection experiments provide constraints on the
$c$ parameters.  Although as discussed above, the limits on the
direct-detection rate constrain additional combinations of Wilson
coefficients and nucleon matrix elements, so far standard SI analyses
have only considered the coefficient $c_+^M$, which is then related to
the WIMP--nucleon cross section by $\sigma_{\chi N}^\text{SI}=
\muN^2|c_+^M|^2/\pi$, with reduced mass $\muN$.  Ideally, to go beyond
this approximation a global correlated analysis of direct-detection
experiments based on either Eq.~\eqref{cross_section_dec} or
Eq.~\eqref{cross_section_dec_min} should be performed in order to
determine limits on all parameters at once, which, however, would
require the consideration of more than one target nucleus in the
analysis.

Barring such a global analysis, one would need to consider slices through the BSM parameter space, e.g.,  in terms of scans over the Wilson coefficients as in Ref.~\cite{Cirigliano:2013zta}.
Such slices through the parameter space could also be organized
in a straightforward extension of present analyses by considering
one nuclear response at a time (this is, setting all but one $c$ to zero),
for instance based on Eq.~\eqref{cross_section_dec_min}, with $4$ $c$ parameters
[which map onto $7$ ($4$) Wilson coefficients for a Dirac (Majorana) WIMP].
This would allow one to set limits on different combinations of Wilson coefficients. 
In particular, due to the role of the two-body responses this kind of analysis
would extend the sensitivity of direct-detection experiments to more new-physics 
couplings than the standard SI single-nucleon cross section studied so far. 
Depending on the sensitivity of the experiment to the $\qq^2$-dependence,
the number of relevant structure factors may be reduced,
and limits could also be obtained for combinations
of the coefficients associated with responses with similar $\qq^2$-tail,
e.g., $\F_\pi$ and $\F_\pi^\theta$.
In that case the one-response-at-a-time analysis could also be
performed based on Eq.~\eqref{cross_section_dec},
which originally depends on $8$ non-independent $c$ coefficients.

In conclusion, we provide a parameterization of the WIMP--nucleus cross section
for general SI scattering, which could be applied to generalize
the extraction of limits from
SI scattering beyond the standard $\sigma_{\chi N}^\text{SI}$ cross
section (corresponding to $c_+^M$), e.g., by similar exclusion plots for the
additional coefficients in the minimal $4$-parameter extension
in Eq.~\eqref{cross_section_dec_min}, or by more sophisticated scans through the BSM parameter space.  
For a xenon target, all necessary structure factors are provided in Secs.~\ref{sec:one-body}
and~\ref{sec:two-body}.

\section{Formalism}
\label{sec:formalism}

We consider a WIMP $\chi$ scattering off a target nucleus $\N$ with momenta assigned as
\beq
\N(p)+\chi(k)\to \N(p')+\chi(k'),
\eeq
and momentum transfer 
\beq
q=k'-k=p-p',\qquad q^2=t,
\eeq
as well as
\beq
P=p+p',\qquad K=k+k'.
\eeq
The rate for the detection of a dark-matter particle $\chi$ scattering elastically off a nucleus with mass number $A$, differential in the three-momentum transfer $\qq$, is then given by
\beq
\label{rate}
\frac{\diff R}{\diff\qq^2}=\frac{\rho M}{\mA\mc}\int_{v_\text{min}}^{v_\text{esc}}\diff^3 v\, |\vv|f(|\vv|)\,\frac{\diff \sigma_{\chi\N}}{\diff \qq^2},
\eeq
where $M$ denotes the (fiducial) mass of the experiment, $\mA$ and $\mc$ the masses of target nucleus and WIMP, respectively, $\sigma_{\chi\N}$ the WIMP--nucleus cross section in the lab frame, $f(|\vv|)$ the normalized velocity distribution of the WIMP, $\rho$ the WIMP density, $v_\text{esc}= 544^{+64}_{-46}\,\text{km}\,\text{s}^{-1}$~\cite{Smith:2006ym} the escape velocity of our galaxy, and 
\begin{align}
v_\text{min}^2&=-t\left[\frac{\sqrt{4\mA^2-t}+\sqrt{4\mc^2-t}}{\sqrt{4\mA^2-t}\sqrt{4\mc^2-t}-t}\right]^2\notag\\
&=\frac{\qq^2}{4\muA^2}+\Order\big(\qq^4\big),\qquad \muA=\frac{\mA\mc}{\mA+\mc},
\end{align}
with $t=-\qq^2$ up to relativistic corrections.
The value for the local WIMP density canonically used in the interpretation of direct-detection experiments is $\rho=0.3\GeV/\text{cm}^3$, although halo-independent methods have been developed that allow one to 
eliminate the astrophysical uncertainties in the comparison of different experiments, see, e.g., Refs.~\cite{Drees:2008bv,Fox:2010bz}.
Alternatively, the detection rate  Eq.~\eqref{rate} is often formulated differential in the recoil energy 
\beq
E_\text{r}=\frac{\qq^2}{2\mA}.
\eeq

The WIMP--nucleus cross section itself combines physics from particle,
hadronic, and nuclear scales. To separate the nuclear contributions,
$\sigma_{\chi\N}$ can be expressed in terms of structure
factors~\cite{Engel:1992bf} 
\beq 
\label{cross_zeta}
\frac{\diff \sigma_{\chi\N}}{\diff\qq^2}=\frac{8 G_F^2}{\vv^2(2J+1)}\Big[S_S(\qq^2)+S_A(\qq^2)\Big], 
\eeq 
where $J$ refers to the spin of the target nucleus,  
$G_F$ denotes the Fermi constant, and $S_S$ and $S_A$ are the structure factors for SI and SD
scattering, respectively. These structure factors are normalized
according to
\begin{align}
\label{SS_SA}
S_S(0)&=\frac{2J+1}{4\pi}\Big|c_0A+c_1 (Z-N)\Big|^2,\notag\\[1mm]
S_A(0)&=\frac{(2J+1)(J+1)}{4\pi J}\notag\\
&\times\Big|(a_0+a_1)\langle\spin_p\rangle+(a_0-a_1)\langle\spin_n\rangle\Big|^2,
\end{align}
with proton and neutron numbers $Z$ and $N$ ($A=Z+N$) and proton/neutron spin expectation values
$\langle \spin_{p/n}\rangle$. 
The constants $c_i$, $a_i$ contain the information about particle and hadronic physics, a relation to be made more precise below. Assuming $c_1=0$, the cross section for SI scattering is often represented in the standard form~\cite{Schumann:2015wfa}
\beq
\label{SI_simp}
\frac{\diff \sigma_{\chi\N}^\text{SI}}{\diff \qq^2}=\frac{\sigma_{\chi N}^\text{SI}}{4\vv^2\mu_N^2}\F_\text{SI}^2(\qq^2),\qquad \muN=\frac{\mN\mc}{\mN+\mc},
\eeq
with nucleon mass $\mN$ and single-nucleon cross section $\sigma_{\chi N}^\text{SI}$. 
The nuclear-physics quantity $\F_\text{SI}(\qq^2)$ is the only remnant of the structure factor,
and is frequently approximated by~\cite{Lewin:1995rx}
\begin{align}
\label{Helm}
\F_\text{SI}^\text{Helm}(\qq^2)&=A\frac{3j_1(|\qq| r_n)}{|\qq| r_n}e^{-\frac{1}{2}\qq^2s^2},\\
r_n^2&=c^2+\frac{7}{3}\pi^2a^2-5s^2,\qquad s=1\,\text{fm},\notag\\
c&=(1.23\,A^{1/3}-0.60)\,\text{fm},\qquad a=0.52\,\text{fm},\notag
\end{align}
whose square is known as Helm form factor.

In the following, we revisit this formalism starting from an effective
Lagrangian for the interaction of the WIMP with Standard-Model fields
presented in Sec.~\ref{sec:lagrangian}.  In Secs.~\ref{sec:scalar}
and~\ref{sec:vector} we discuss the relevant nucleon couplings and
finally in Sec.~\ref{sec:structure} we derive a generalized
decomposition for SI scattering that includes two-body currents and
the nuclear $\Phi''$ response.

\subsection{Lagrangian and Wilson coefficients}
\label{sec:lagrangian}

We consider the following dimension-$6$ and -$7$ effective
Lagrangian for the interaction of a spin-$1/2$ WIMP with quark and gluon fields
\begin{align}
\label{Lagr}
\mathcal{L}_{\chi}&=\mathcal{L}_{\chi}^{(6)}+\mathcal{L}_{\chi}^{(7)},\notag\\
\mathcal{L}_{\chi}^{(6)}&=\frac{1}{\Lambda^2}\sum_{q}\Big[C_q^{VV}\bar\chi\gamma^\mu\chi \,\bar q\gamma_\mu q
+C_q^{AA}\bar\chi\gamma^\mu\gamma_5\chi\, \bar q\gamma_\mu\gamma_5 q\Big],\notag\\
\mathcal{L}_{\chi}^{(7)}&=
\frac{1}{\Lambda^3}\sum_{q} C_{q}^{SS}\bar \chi \chi \,m_q\bar q q
+\frac{1}{\Lambda^3}C'^S_{g}\bar \chi\chi\, \alpha_sG_{\mu\nu}^aG^{\mu\nu}_a\notag\\
&=\frac{1}{\Lambda^3}\sum_{q} \Big(C_{q}^{SS}+\frac{8\pi}{9}C'^S_{g}\Big)\bar \chi \chi \,m_q\bar q q\notag\\
&\qquad-\frac{8\pi}{9}\frac{1}{\Lambda^3}C'^S_{g}\bar \chi\chi\, \theta^\mu_\mu,
\end{align}
where $\chi$ is assumed to be a Standard-Model singlet, the quark
masses $m_q$ have been included to make the scalar operator
renormalization-group invariant, and the Wilson coefficients $C_i$
parameterize the BSM physics associated with the scale $\Lambda$.
The effective Lagrangian is defined at the hadronic scale, with the quark sum
extending over $q=u,d,s$, after the heavy quarks have been integrated out
and their effect has been absorbed into a redefinition of the gluon coefficient $C_g^S$, 
see Eq.~\eqref{shifman_trick}. In the second formulation of the dimension-$7$ Lagrangian
the gluon term has been replaced in favor of the trace of the QCD energy-momentum tensor
$\theta^\mu_\mu$.
Equation~\eqref{Lagr} includes the leading operators relevant for
coherent WIMP--nucleus scattering, vector and scalar channels, but
also retains the axial-vector operator to facilitate the comparison to
the SD case. The WIMP could either be a Dirac or Majorana particle,
with $C_q^{VV}=0$ in the latter case.
At dimension $8$, there are
spin-$2$ operators that can become relevant for the SI scattering of
heavy WIMPs~\cite{Hill:2014yxa}, but their inclusion will be left for
future work. Similarly, the operator basis changes for different 
quantum numbers of the WIMP~\cite{Goodman:2010ku,Hill:2014yxa}.

Throughout this work we follow the chiral counting formulated in
Refs.~\cite{Cirigliano:2012pq,Hoferichter:2015ipa} to organize the
calculation.  In particular, this implies that momentum corrections to
the one-body matrix elements occurring in Eq.~\eqref{Lagr} enter at
the same order as the leading two-body contributions, at third order
in ChEFT~\cite{Hoferichter:2015ipa}.  The nucleon matrix elements of
the operators listed in Eq.~\eqref{Lagr} involve a combination of
Wilson coefficients and nucleon couplings. In the next sections, we
spell out these combinations, closely following the notation
introduced in Ref.~\cite{Hoferichter:2015ipa}.

\subsection{Scalar couplings}
\label{sec:scalar}

For the scalar channel in Eq.~\eqref{Lagr} we need the following coupling to the nucleon ($N = n$ or $p$)
\beq
\label{fN_scalar}
f_N(t)=\frac{m_N}{\Lambda^3}\bigg(\sum_{q=u,d,s}C^{SS}_{q}f_q^N(t)-12\pi f^N_Q(t)C'^S_{g}\bigg),
\eeq
where the nucleon scalar form factors are defined as
\beq
\mN f_q^N(t)=\langle N(p')|m_q\bar q q|N(p)\rangle.
\eeq
The form factors for the heavy quarks $f_Q^N(t)$ appear together with the modified gluon Wilson coefficient
\beq
\label{shifman_trick}
C'^S_{g}=C^S_{g}-\frac{1}{12\pi}\sum_{Q=c,b,t}C^{SS}_{Q}
\eeq
after integrating out their effect by means of the trace anomaly of the energy-momentum tensor $\theta_{\mu\nu}$~\cite{Shifman:1978zn},
which also produces
\begin{align}
\label{fQ}
f_Q^N(t)&=\frac{2}{27}\bigg(\frac{\theta_0^N(t)}{\mN}-\sum_{q=u,d,s}f_q^N(t)\bigg),\notag\\
\theta_0^N(t)&=\langle N(p')|\theta^\mu_\mu|N(p)\rangle.
\end{align}
It should be noted that this procedure is accurate at
$\Order(\alpha_s)$, which may not be sufficient for the $c$ quark, see
Refs.~\cite{Kryjevski:2003mh,Vecchi:2013iza,Hill:2014yxa} for a study
of higher orders in $\alpha_s$.

We begin with the discussion of Eq.~\eqref{fN_scalar} at vanishing
momentum transfer, in which case the form factors simply reduce to the
scalar couplings of the nucleon.  Based on $SU(2)$ chiral perturbation
theory (ChPT), it can be shown that the couplings to $u$ and $d$
quarks only depend on the value of the pion--nucleon $\sigma$-term
$\sigma_{\pi N}$, while isospin-breaking corrections are fully
determined by the same low-energy constant that governs the strong
contribution to the proton--neutron mass
difference~\cite{Crivellin:2013ipa}. Combining dispersive
techniques~\cite{Hoferichter:2015hva} with precision data for the
pion--nucleon scattering lengths extracted from pionic
atoms~\cite{Baru:2010xn,Baru:2011bw} leads to the
phenomenological values~\cite{Hoferichter:2015dsa} for the light-quark
couplings quoted in the first line of
Table~\ref{tab:scalar_ud_couplings}.  More recently, lattice
calculations at physical quark masses have produced significantly lower
values for $\sigma_{\pi N}$~\cite{Durr:2015dna,Yang:2015uis,Abdel-Rehim:2016won,Bali:2016lvx},
which translates to the $3\sigma$ tension in the scalar couplings
shown in Table~\ref{tab:scalar_ud_couplings}. This tension between
phenomenology and lattice~\cite{Hoferichter:2016ocj} currently
constitutes the largest uncertainty in the $u$ and $d$ couplings.

\begin{table}[t]
\centering
\renewcommand{\arraystretch}{1.3}
\begin{tabular}{ccccc}
\toprule
$f^p_u$ & $f^n_u$ & $f^p_d$ & $f^n_d$ & Ref.\\
$20.8(1.5)$ & $18.9(1.4)$ & $41.1(2.8)$ & $45.1(2.7)$ & \cite{Hoferichter:2015dsa} \\
$13.9(1.8)$ & $11.6(1.7)$ & $25.3(3.7)$ & $30.2(3.8)$ & \cite{Durr:2015dna} \\
\botrule
\end{tabular}
\renewcommand{\arraystretch}{1.0}
\caption{Scalar $u$ and $d$ couplings of the nucleon, in units of $10^{-3}$.}
\label{tab:scalar_ud_couplings}
\end{table}

In contrast to the $u$ and $d$ quarks, a determination of the scalar
coupling to the $s$ quark from phenomenology requires the use of
$SU(3)$ relations, whose convergence properties make reliable
uncertainty estimates difficult.  For this reason, in
Table~\ref{tab:scalar_s_couplings} we only quote the values obtained
by recent lattice calculations, together with the average from
Ref.~\cite{Junnarkar:2013ac} of previous lattice results. In
particular, we assume isospin symmetry $f_q^p=f_q^n$ for $q=s,c,b,t$.
Finally, Ref.~\cite{Abdel-Rehim:2016won} also provides a value for the
$c$ coupling, $f_c^N=0.085(22)$, to be compared with $f_Q^N=0.068(1)$
as extracted from the same reference based on Eq.~\eqref{fQ} (with
$\theta_0^N(0)=\mN$). Within uncertainties, the direct determination
from lattice QCD thus agrees with the result extracted by means of the
trace anomaly at $\Order(\alpha_s)$.

\begin{table}[t]
\centering
\renewcommand{\arraystretch}{1.3}
\begin{tabular}{lccccc}
\toprule
$f^N_s$ & $113(60)$ & $34(7)$ & $44(9)$ & $37(13)$ & $43(11)$\\
Ref. & \cite{Durr:2015dna} & \cite{Yang:2015uis} & \cite{Abdel-Rehim:2016won} & \cite{Bali:2016lvx} & \cite{Junnarkar:2013ac}\\
\botrule
\end{tabular}
\renewcommand{\arraystretch}{1.0}
\caption{Scalar $s$ coupling of the nucleon, in units of $10^{-3}$.}
\label{tab:scalar_s_couplings}
\end{table}

Next, we turn to the finite-momentum-transfer corrections to
$f_N(0)\equiv f_N$.\footnote{Here and below, couplings without
argument are understood to be evaluated at $t=0$.} To the order we
are working in ChEFT, it is generally sufficient to keep the radius
corrections, i.e., the first order in the expansion around $t=0$.
However, the strong $\pi\pi$ rescattering in the isospin-$0$ $\pi\pi$
$S$-wave makes the leading-loop ChPT prediction for the slope of the
scalar form factor of the nucleon at $t=0$~\cite{Gasser:1990ap}
\beq
\label{sigma_dot}
\dot\sigma\big|_\text{ChPT}=\frac{5\gA^2\mpi}{256\pi\Fpi^2}=0.17\GeV^{-1},
\eeq
underestimate the true result by nearly a factor of $2$ ($\gA=1.2723(23)$ and $\Fpi=92.2(2)\MeV$ are taken from Ref.~\cite{Agashe:2014kda}). For this reason, we make use of the updated dispersive analysis from Refs.~\cite{Hoferichter:2012wf,Ditsche:2012fv} and use
\beq
\dot\sigma=0.27(1)\GeV^{-1}.
\eeq
Retaining the leading isospin-breaking effect, this correction amounts to the replacement
\begin{align}
f_u^N(t)&\to f_u^N+\frac{1-\xi}{2\mN}\dot\sigma t,\qquad \xi=\frac{\md-\muu}{\md+\muu}=0.37(3),\notag\\
f_d^N(t)&\to f_d^N+\frac{1+\xi}{2\mN}\dot\sigma t,
\end{align}
where we have used $\muu/\md=0.46(3)$~\cite{Aoki:2013ldr}.

In analogy to Eq.~\eqref{sigma_dot}, there is a parameter-free
prediction from leading-loop $SU(3)$ ChPT for the slope of the
strangeness radius~\cite{Cirigliano:2012pq}
\begin{align}
\dot\sigma_s\big|_\text{ChPT}&=\frac{5\gA^2}{256\pi\Fpi^2}\Big(M_K^2-\frac{1}{2}\mpi^2\Big)\frac{1}{3}\Bigg\{\frac{4}{3M_\eta}\bigg(\frac{1-4\alpha}{\sqrt{3}}\bigg)^2\notag\\
&+\frac{1}{M_K}\Bigg[3(1-2\alpha)^2+\bigg(\frac{1+2\alpha}{\sqrt{3}}\bigg)^2\Bigg]\Bigg\}\notag\\
&=0.24\GeV^{-1},
\end{align}
where $\alpha=F/(D+F)$ parameterizes the leading $SU(3)$ couplings. Numerically, we use $F/D=0.57$ as extracted from semileptonic hyperon decays~\cite{Ratcliffe:1995fk,Yamanishi:2007zza}, which together with the $SU(2)$ constraint $D+F=\gA$ implies
\beq
D=0.81,\qquad F=0.46,\qquad \alpha=0.36.
\eeq
However, such $SU(3)$ leading-loop low-energy theorems are known to be sensitive to higher-order corrections~\cite{Hemmert:1998pi,Hammer:2002ei}. Therefore, we also considered the coupled-channel dispersive analysis~\cite{Hoferichter:2012wf}, which in principle provides not only a prediction for $\dot\sigma$ but also for $\dot\sigma_s$. Unfortunately, convergence of the dispersive integrals is much slower for the slope of the strangeness form factor, although the resulting values are not too far from the chiral prediction. All in all, the spread observed in both methods would be covered by a range
\beq
\dot\sigma_s=0.3(2)\GeV^{-1},
\eeq
leading to
\beq
f_s^N(t)\to f_s^N+\frac{\dot\sigma_s}{\mN} t.
\eeq
In view of the substantial uncertainties already encountered in the
strangeness form factor, we do not make an attempt to quantify radius
corrections for the heavy quarks. 
The leading chiral result, however, can be reconstructed by means of Eq.~\eqref{fQ}
and 
\beq
\theta_0^N(t)=\mN-\frac{13\gA^2\mpi}{128\pi\Fpi^2} t+\Order\big(t^2\big). 
\eeq

Taking everything together, we arrive at the following decomposition
of the combination of Wilson coefficients and nucleon form factors
relevant for the scalar channel
\begin{align}
\label{scalar_couplings}
f_N(t)&=f_N+t \dot f_N+\Order\big(t^2\big),\\
f_N&=\frac{m_N}{\Lambda^3}\bigg(\sum_{q=u,d,s}C^{SS}_{q}f_q^N-12\pi f^N_QC'^S_{g}\bigg),\notag\\
\dot f_N&=\frac{1}{\Lambda^3}\bigg(C^{SS}_{u}\frac{1-\xi}{2}\dot\sigma+C^{SS}_{d}\frac{1+\xi}{2}\dot\sigma+C^{SS}_{s}\dot\sigma_s\bigg).\notag
\end{align}

For the scalar two-body matrix element we also need the couplings to the pion
\begin{align}
\label{fpi}
f_\pi&=\frac{\mpi}{\Lambda^3}\sum_{q=u,d} \Big(C_{q}^{SS}+\frac{8\pi}{9}C'^S_{g}\Big)f_q^\pi,\notag\\
f_\pi^\theta&=-\frac{\mpi}{\Lambda^3}\frac{8\pi}{9}C'^S_{g},
\end{align}
with
\begin{align}
f_u^\pi&=\frac{\muu}{\muu+\md}=\frac{1}{2}\big(1-\xi\big)=0.32(2),\notag\\
f_d^\pi&=\frac{\md}{\muu+\md}=\frac{1}{2}\big(1+\xi\big)=0.68(2).
\end{align}
In Eq.~\eqref{fpi} we introduced a factor $\mpi$ in analogy to the
scalar coupling to the nucleon, Eq.~\eqref{fN_scalar}. 
The necessity of defining two pion couplings, $f_\pi$ and $f_\pi^\theta$ in Eq.~\eqref{fpi}, 
traces back to the fact that the couplings of the scalar current $m_q\bar q q$ and the trace anomaly
of the energy-momentum tensor $\theta^\mu_\mu$ to the pion differ qualitatively: while the former is constant up to higher-order corrections, the latter 
becomes momentum dependent and therefore produces a different nuclear structure factor.

\subsection{Vector and axial-vector couplings}
\label{sec:vector}

In the vector channel there are two sets of couplings to the nucleon
\beq
f_i^{V,N}(t)=\frac{1}{\Lambda^2}\sum_{q=u,d,s}C^{VV}_{q}F_i^{q,N}(t),
\eeq
with $i=1,2$ related to the Dirac and Pauli terms, respectively, in the decomposition of the nucleon form factors of the electromagnetic current. A decomposition analogous to Eq.~\eqref{scalar_couplings} is given by
\begin{align}
f_1^{V,N}(t)&=f_1^{V,N}+t\dot f_1^{V,N} +\Order\big(t^2\big),\notag\\[1mm]
f_2^{V,N}(t)&=f_2^{V,N}+\Order(t).
\end{align}
Since the matrix element of the Pauli form factor vanishes at zero
momentum transfer, the leading term in $f_2^{V,N}(t)$ is
sufficient. Assuming isospin symmetry (for corrections see
Ref.~\cite{Kubis:2006cy}), these couplings expressed in terms of
nucleon radii and anomalous magnetic moments
become~\cite{Hoferichter:2015ipa}
\begin{align}
\label{vector_couplings}
f_1^{V,p}&=\frac{1}{\Lambda^2}\Big(2C^{VV}_{u}+C^{VV}_{d}\Big),\\
f_2^{V,p}&=\frac{1}{\Lambda^2}\bigg[\Big(2C^{VV}_{u}+C^{VV}_{d}\Big)\kappa_p+\Big(C^{VV}_{u}+2C^{VV}_{d}\Big)\kappa_n\notag\\
&+\Big(C^{VV}_{u}+C^{VV}_{d}+C^{VV}_{s}\Big)\kappa_N^s\bigg],\notag\\
\dot f_1^{V,p}&=\frac{1}{\Lambda^2}\bigg[\Big(2C^{VV}_{u}+C^{VV}_{d}\Big)\bigg(\frac{\langle r_E^2\rangle^p}{6}-\frac{\kappa_p}{4\mpp^2}\bigg)\notag\\
&+\Big(C^{VV}_{u}+2C^{VV}_{d}\Big)\bigg(\frac{\langle r_E^2\rangle^n}{6}-\frac{\kappa_n}{4\mn^2}\bigg)\notag\\
&+\Big(C^{VV}_{u}+C^{VV}_{d}+C^{VV}_{s}\Big)\bigg(\frac{\langle r_{E,s}^2\rangle^N}{6}-\frac{\kappa_N^s}{4\mN^2}\bigg)\bigg],\notag
\end{align}
and $u\leftrightarrow d$ for the neutron couplings.  Numerical values
for the nucleon radii and anomalous magnetic moments are collected in
Table~\ref{tab:vector_coupling}.

\begin{table}[t]
\centering
\renewcommand{\arraystretch}{1.3}
\begin{tabular}{ccc}
\toprule
$\kappa_p$ & $\kappa_n$ & $\kappa_N^s$\\
$1.792847356(23)$ & $-1.91304272(45)$ & $-0.26(26)$\\
$\langle r_E^2\rangle^p$ & $\langle r_E^2\rangle^n$ & $\langle r_{E,s}^2\rangle^N$\\
$0.7071(7)\fm^2$ & $-0.1161(22)\fm^2$ & $-0.06(4)\fm^2$\\
\botrule
\end{tabular}
\renewcommand{\arraystretch}{1.0}
\caption{Nucleon radii and anomalous magnetic moments.
The values of $\kappa_p$, $\kappa_n$, and $\langle r_E^2\rangle^n$ are
taken from Ref.~\cite{Agashe:2014kda}, $\langle r_E^2\rangle^p$ from
Ref.~\cite{Antognini:1900ns}, and $\kappa_N^s$ as well as $\langle
r_{E,s}^2\rangle^N$ from a global analysis of parity-violating
asymmetry data~\cite{Gonzalez-Jimenez:2014bia}.  Note that the latter
two are strongly correlated, with a correlation coefficient $0.87$.}
\label{tab:vector_coupling}
\end{table}

For completeness, we also quote the analogous decomposition for the
axial-vector channel appearing in Eq.~\eqref{Lagr}. In this case, one
needs the combinations $g_A^N(t)$ and $g_P^N(t)$ with
\begin{align}
\label{axial_couplings}
g_A^N(t)&=g_A^N+t\dot g_A^N+\Order\big(t^2\big),\notag\\
g_A^N&=\frac{1}{\Lambda^2}\bigg[\pm\frac{\gA}{2}\Big(C^{AA}_u-C^{AA}_d\Big)\notag\\
&+\frac{3F-D}{6}\Big(C^{AA}_u+C^{AA}_d-2C^{AA}_s\Big)\notag\\
&+\frac{\Delta\!\Sigma^N}{3}\Big(C^{AA}_u+C^{AA}_d+C^{AA}_s\Big)\bigg],\notag\\
\dot g_A^N&=\pm\frac{g_A}{\Lambda^2}\Big(C^{AA}_u-C^{AA}_d\Big)\frac{1}{M_A^2},\notag\\
g_P^N(t)&=-\frac{4\mN^2}{\Lambda^2}\bigg[\pm\frac{\gA}{2}\Big(C^{AA}_u-C^{AA}_d\Big)\frac{1}{t-\mpi^2}\notag\\
&+\frac{3F-D}{6}\Big(C^{AA}_u+C^{AA}_d-2C^{AA}_s\Big)\frac{1}{t-\meta^2}\bigg],
\end{align}
where the upper/lower sign refers to proton/neutron and the small
$\eta$ contribution of the last line above is generally neglected in
SD analyses. These relations involve the nucleon spin matrix elements
$\Delta q^N = \langle N|\bar q \gamma_\mu\gamma_5 q|N \rangle /
\langle N|\gamma_\mu\gamma_5|N\rangle$, for which we have assumed
isospin symmetry and already used the combinations
\begin{align}
\gA&=\Delta u^p-\Delta d^p=\Delta d^n-\Delta u^n, \notag\\
3F-D&=\Delta u^N+\Delta d^N-2\Delta s^N,\notag\\
\Delta\!\Sigma^N&=\Delta u^N+\Delta d^N+\Delta s^N.
\end{align}
Due to the axial anomaly, the singlet combination $\Delta\!\Sigma^N$
cannot be analyzed in $SU(3)$ ChPT, as effects related to the $\eta'$
will play a role. However, it can be extracted from the spin structure
function of the nucleon, which, at $Q^2=5\GeV^2$ and to order
$\Order(\alpha_s^2)$, produces
$\Delta\!\Sigma^N=0.330(39)$~\cite{Airapetian:2006vy}.  Further, the
dominant radius correction occurring in the isovector contribution in
Eq.~\eqref{axial_couplings} has been included by a dipole ansatz with
mass parameter $M_A$ around
$1\GeV$~\cite{Bernard:2001rs,Schindler:2006it}, while the pseudoscalar
poles in $g_P^N(t)$ prevent a Taylor expansion in $t$.

\subsection{Structure factors}
\label{sec:structure}

For the definition of the nuclear structure factors we first consider
the matching of the one-body operators obtained in ChEFT above onto
the NREFT basis of
Refs.~\cite{Fitzpatrick:2012ix,Anand:2013yka}.
This produces the matrix elements
\begin{align}
\label{matching}
\M_{1,\text{NR}}^{SS}&= \Op_1 f_N(t),\notag\\
\M_{1,\text{NR}}^{VV}&=\Op_1\Big(f_1^{V,N}(t)+\frac{t}{4\mN^2}f_2^{V,N}(t)\Big)\notag\\
&+\frac{1}{\mN}\Op_3 f_2^{V,N}(t),\notag\\
\M_{1,\text{NR}}^{AA}&=
-4\Op_4 g_A^N(t)+\frac{1}{\mN^2}\Op_6 g_P^N(t),
\end{align}
where we have dropped the nucleon and WIMP spinors.\footnote{For details see Ref.~\cite{Hoferichter:2015ipa}. This matching 
is performed at tree level and hence
does not include effects from operator evolution, which could be generated when running the ChEFT operators down to nuclear scales.}
The NREFT operators $\Op_i$ are defined by
\begin{align}
\label{Op_Wick}
\Op_1&=\mathds{1}, & \Op_3&=i\spin_N\cdot (\qq\times\vvp),\notag\\ 
\Op_4&=\spin_\chi\cdot \spin_N,  & \Op_6&=\spin_\chi\cdot \qq\,\spin_N\cdot \qq,
\end{align}
with spins $\spin=\sig/2$ and velocity
\beq
\label{def_vel}
\vvp=\frac{\KK}{2\mc}-\frac{\PP}{2\mN}.
\eeq
The combination of the different operators in Eq.~\eqref{matching}
demonstrates how QCD constraints impose relations between the NREFT
operators: for the axial-vector channel it is a fixed combination of
$\Op_4$ and $\Op_6$ that contributes, while the same coefficient
$f_2^{V,N}(t)$ that multiplies $\Op_3$ also appears as a
momentum-dependent correction to $\Op_1$.

In Eq.~\eqref{matching} we only retained those channels that generate
coherent or quasi-coherent nuclear responses, compared to the full
list studied in Ref.~\cite{Hoferichter:2015ipa}.  These coherent and
quasi-coherent responses are denoted as $M$ and $\Phi''$ in
Refs.~\cite{Fitzpatrick:2012ix,Anand:2013yka}, and
are only a subset of the six different nuclear responses generated by
the NREFT operators, which also include the $\Sigma'$, $\Sigma''$,
$\Delta$, and $\tilde\Phi'$ responses.  For example, $M$ governs
standard SI scattering, and it is a combination of $\Sigma'$ and
$\Sigma''$ that enters in SD scattering.

Beyond the single-nucleon sector, NREFT operators that involve
$\vv^\perp$ can be decomposed into two
parts~\cite{Fitzpatrick:2012ix,Anand:2013yka}.  First, there are terms
proportional to the relative WIMP velocity with respect to the
center-of-mass of the nucleus
\beq
\label{vperpT}
\vv^\perp_T=\frac{\KK}{2\mc}-\frac{1}{A}\sum_{i=1}^A\frac{\PP_i}{2\mN},
\eeq
where $\PP_i=\pp_i+\pp'_i$ is the sum of the initial and final nucleon momenta.
These terms are effectively suppressed by the WIMP velocity with
respect to the target $|\vv^\perp_T|\approx 10^{-3}$ and will thus be
neglected in the following.  Second, $\vv^\perp$ also produces
contributions involving the velocity operator of the nucleon, which
are part of the $\Delta$, $\tilde\Phi'$, and $\Phi''$ responses and
come with a milder suppression factor $|\qq|/\mN$.  This is the case
for the $\Op_3$ contribution kept in Eq.~\eqref{matching}, which
generates a $\Phi''$ response.  In the end, for coherent SI scattering
only scalar and vector interactions remain, and the fact that the
$\Phi''$ response is due only to the vector operator could serve as a tool
to discriminate between these two channels.

\begin{figure}[t] 
\centering
\includegraphics[width=\columnwidth,clip]{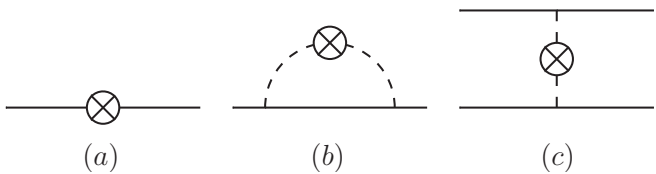}
\caption{Diagrams for WIMP--nucleon interactions in ChEFT. Solid (dashed)
lines denote nucleons (pions) and crosses the coupling to the WIMP
current.  Diagram $(a)$ represents a leading one-body term, $(b)$ a
radius correction, and $(c)$ a two-body current.}
\label{fig:diagrams}
\end{figure}

Apart from the one-body operators and the momentum corrections as
summarized in Secs.~\ref{sec:scalar} and~\ref{sec:vector}, there are
two-body currents at the same order in ChEFT, see
Fig.~\ref{fig:diagrams}. The corresponding NR
amplitudes take the form
\begin{align}
\label{scalar_two_body}
\M_{2,\text{NR}}^{SS}&=-\bigg(\frac{\gA}{2\Fpi}\bigg)^2f_\pi\mpi \frac{\ttau_1\cdot\ttau_2\,\sig_1\cdot\qq_1\,\sig_2\cdot\qq_2}{\big(\qq_1^2+\mpi^2\big)\big(\qq_2^2+\mpi^2\big)},\notag\\
\M_{2,\text{NR}}^{\theta}&=\frac{4\mpi^2-2\qq_1\cdot \qq_2}{\mpi^2}\frac{f_\pi^\theta}{f_\pi}\M_{2,\text{NR}}^{SS},
\end{align}
where $f_\pi$ and $f_\pi^\theta$ are defined in Eq.~\eqref{fpi},
$\sig_i$ and $\ttau_i$ denote
the spin and isospin Pauli matrices of nucleon $i$, respectively, and
$\qq_i=\pp_i'-\pp_i$. Diagrammatically, these amplitudes represent the coupling of 
the WIMP to the pion in flight via a scalar current and by means of the QCD trace anomaly $\theta^\mu_\mu$.
The other two-body currents identified in
Ref.~\cite{Hoferichter:2015ipa} in general involve isospin operators
$[\ttau_1\times\ttau_2]^3$ as well as spin structures that, after
summing over spins, make the diagrams vanish.  The only remaining
contribution is the exchange diagram from the axial-vector--vector
channel, whose isospin structure becomes $\tau_1^3-\tau_2^3$, only
allowing for an isovector coherent enhancement suppressed by $(N-Z)/A$
with respect to the scalar two-body current.  In addition, this
two-body current is linear in $\spin_\chi$ and does not interfere with
$\Op_1$~\cite{Fitzpatrick:2012ix}.  Other contributions such as the
vector--vector two-body current also show isovector coherent
enhancement only, and are further suppressed in the ChEFT
expansion~\cite{Hoferichter:2015ipa}.  For these reasons, we restrict
our analysis to the contributions given by
Eq.~\eqref{scalar_two_body}.
It is the presence of the $\qq_1\cdot \qq_2$ term
in the relation between quark-mass and trace-anomaly couplings
that necessitates the definition of two structure factors:
for a constant term, the $\theta^\mu_\mu$ contribution
could be absorbed into a redefinition of $f_\pi$,
similarly to $f_N$ in the case of the nucleon coupling [see Eqs.~\eqref{fN_scalar} and~\eqref{fQ}].

In this context, several comments on the role of two-body operators
are in order. First, the hierarchy of diagrams shown in
Fig.~\ref{fig:diagrams} assumes the ChEFT counting originally proposed
by Weinberg~\cite{Weinberg:1990rz,Weinberg:1991um}.  In this counting,
the coupling of the scalar current to $(N^\dagger N)^2$ contact
operators is suppressed by two orders in the chiral expansion.  Due to
the limitations of Weinberg counting, this suppression might be less
pronounced in practice, as indicated, e.g., by KSW
counting~\cite{Kaplan:1998tg,Kaplan:1998we} or by general arguments
related to the short-range behavior of nucleon--nucleon wave
functions~\cite{Valderrama:2014vra}.  The role of such contact
operators at heavy pion masses has been studied in
Ref.~\cite{Beane:2013kca} using lattice QCD, and calculations at or close
to the physical pion mass would allow for a check of the ChEFT counting
employed here.

Second, while diagram $(a)$ corresponds directly to an NREFT operator
from Refs.~\cite{Fitzpatrick:2012ix,Anand:2013yka}, the mapping of
diagrams $(b)$ and $(c)$ would proceed in an indirect way. The radius
corrections $(b)$ are represented by $\qq$-dependent prefactors of the
$\Op_i$, see Eq.~\eqref{matching}. The two-body contributions $(c)$
could be modeled as effective one-body operators, if summed over the
second nucleon with respect to a given reference state, symbolically
written as $\langle N^\dagger N\rangle N^\dagger N$, so that the
effective one-body operator would become density and state dependent.
Such a normal-ordering approximation with respect to a Fermi gas was
used in the context of SD
scattering~\cite{Menendez:2012tm,Klos:2013rwa}.  However, in this work
we perform a full calculation in harmonic-oscillator basis states, see
Sec.~\ref{sec:two-body}. It is the explicit calculation of all
diagrams $(a)$--$(c)$ within ChEFT, instead of a parameterization in
terms of effective one-body operators, that allows one to relate the
coefficients of the nuclear structure factors to nucleon form factors
and new-physics parameters.

For the construction of suitable nuclear structure factors for
generalized SI scattering, we first turn to the SD case. Here the
result in Eq.~\eqref{axial_couplings} shows that once the $\eta$
contribution to $g_P^N(t)$ is neglected, only two independent
combinations of Wilson coefficients remain, which can be conveniently
identified with the coefficients introduced in Eq.~\eqref{SS_SA}
\begin{align}
\label{a0a1_matching}
a_0&=\frac{\zeta}{2\sqrt{2}G_F\Lambda^2}\notag\\
&\times\bigg[\Big(C^{AA}_u+C^{AA}_d\Big)\big(\Delta u^N+\Delta d^N)+2C^{AA}_s\Delta s^N\bigg]\notag\\
&=\frac{\zeta}{6\sqrt{2}G_F\Lambda^2}\bigg[(3F-D)\Big(C^{AA}_u+C^{AA}_d-2C^{AA}_s\Big)\notag\\
&+2\Delta\!\Sigma\Big(C^{AA}_u+C^{AA}_d+C^{AA}_s\Big)\bigg],\notag\\
a_1&=\frac{\zeta}{2\sqrt{2}G_F\Lambda^2}\Big(C^{AA}_u-C^{AA}_d\Big)\big(\Delta u^p-\Delta d^p)\notag\\
&=\frac{\zeta\gA}{2\sqrt{2}G_F\Lambda^2}\Big(C^{AA}_u-C^{AA}_d\Big),
\end{align}
where $\zeta=1 (2)$ for a Dirac (Majorana) spin-$1/2$ WIMP.
The structure factor can therefore be decomposed as
\beq
\label{SA_decomposition}
S_A(\qq^2)=a_0^2S_{00}(\qq^2)+a_0a_1 S_{01}(\qq^2)+a_1^2S_{11}(\qq^2),
\eeq
or, in terms of so-called proton-only and neutron-only structure factors,
\begin{align}
S_A^p(\qq^2)&=S_{00}(\qq^2)+ S_{01}(\qq^2)+S_{11}(\qq^2),\notag\\
S_A^n(\qq^2)&=S_{00}(\qq^2)- S_{01}(\qq^2)+S_{11}(\qq^2).
\end{align}
Since both the momentum corrections in Eq.~\eqref{axial_couplings} and
the leading two-body currents~\cite{Klos:2013rwa} also depend on $a_0$
and $a_1$ only, this implies that the definition of the structure
factors Eq.~\eqref{SA_decomposition} remains applicable even once such
corrections are included. In fact, in a normal-ordering approximation
the effect from two-body currents amounts to a shift $a_1\to
a_1(1+\Delta a_1)$, with $\Delta a_1$ predicted from ChEFT. The
connection between experimental limits for the direct-detection rate
and the Wilson coefficients therefore still proceeds by means of
Eq.~\eqref{a0a1_matching}.

Our aim is to find a similar decomposition for SI scattering.  More
precisely, we wish to formulate a set of structure factors that
captures the leading corrections, taking into account both the ChEFT
expansion and coherence effects in the nucleus, in particular
including both one- and two-body operators.

As a first step towards the construction of generalized SI structure
factors, we again identify the couplings at vanishing momentum
transfer. In this limit we obtain
\begin{align}
c_0&=\frac{\zeta}{4\sqrt{2}G_F}\big(f_p+f_n+f_1^{V,p}+f_1^{V,n}\big),\notag\\
c_1&=\frac{\zeta}{4\sqrt{2}G_F}\big(f_p-f_n+f_1^{V,p}-f_1^{V,n}\big).
\end{align}
Indeed, for $f_p=f_n=f_N$, $f_1^{V,p}=f_1^{V,n}=f_1^{V,N}$ the single-nucleon cross section at threshold becomes
\beq
\label{sigma_Wilson_coefficient}
\sigma_{\chi N}^\text{SI}=\frac{\zeta^2\muN^2}{\pi}\Big|f_N+f_1^{V,N}\Big|^2,
\eeq
leading to the simplification anticipated in Eq.~\eqref{SI_simp}.
Limits for $\sigma_{\chi N}^\text{SI}$ should therefore be interpreted
as limits on the combination of Wilson coefficients given by
$f_N+f_1^{V,N}$, under the assumption that proton and neutron
couplings are identical.

Based on the previous discussion we propose the following decomposition
for the WIMP--nucleus differential cross section
\begin{align}
\label{structure_factors_def}
\frac{\diff \sigma_{\chi\N}^\text{SI}}{\diff \qq^2}&=\frac{\zeta^2}{4\pi\vv^2}\Big|f_+^M(\qq^2)\F_+^M(\qq^2)+f_-^M(\qq^2)\F_-^M(\qq^2)\notag\\
&+\frac{\qq^2}{2\mN^2}\big[f_+^{\Phi''}\F_+^{\Phi''}(\qq^2)+f_-^{\Phi''}\F_-^{\Phi''}(\qq^2)\big]\notag\\
&+f_\pi \F_\pi(\qq^2)+f_\pi^\theta \F_\pi^\theta(\qq^2)\Big|^2,
\end{align}
where
\begin{align}
\label{form_factors}
f_\pm^M(\qq^2)&=\frac{1}{2}\bigg[f_p\pm f_n-\qq^2\Big(\dot f_p\pm \dot f_n\Big)\notag\\
&+f_1^{V,p}\pm f_1^{V,n}-\qq^2\Big(\dot f_1^{V,p}\pm \dot f_1^{V,n}\Big)\notag\\
&-\frac{\qq^2}{4\mN^2}\Big(f_2^{V,p}\pm f_2^{V,n}\Big)\bigg],\notag\\
f_\pm^{\Phi''}&=\frac{1}{2}\Big(f_2^{V,p}\pm f_2^{V,n}\Big).
\end{align}
The nuclear $M$ responses in Eq.~\eqref{structure_factors_def} are normalized to
\beq
\F_+^M(0)=A,\qquad \F_-^M(0)=Z-N,
\eeq
so that $\F_+^M(\qq^2)$ coincides with the standard SI response
$\F_\text{SI}(\qq^2)$ in Eq.~\eqref{SI_simp}, and at vanishing
momentum transfer is given by the combination of couplings that
determines the single-nucleon cross section, see
Eq.~\eqref{sigma_Wilson_coefficient}.  In addition, $\F_-^M(\qq^2)$
provides the corresponding isovector piece, $\F_\pm^{\Phi''}(\qq^2)$
is generated by the $\Op_3$ operator in the vector channel, and
$\F_\pi(\qq^2)$ and $\F_\pi^\theta(\qq^2)$ represent the two-body-current contributions.
It is this decomposition in Eq.~\eqref{structure_factors_def} that
underlies the analysis strategy discussed in Sec.~\ref{sec:overview}.
The nuclear response functions $\F_\pm^M(\qq^2)$,
$\F_\pm^{\Phi''}(\qq^2)$, $\F_\pi(\qq^2)$, and $\F_\pi^\theta(\qq^2)$ are the subject of
Secs.~\ref{sec:one-body} and \ref{sec:two-body}, where simple
parameterizations are provided.

In Eq.~\eqref{structure_factors_def} we used the interference pattern
for the one-body pieces found in
Refs.~\cite{Fitzpatrick:2012ix,Anand:2013yka} for $L=0$ multipoles,
and extended it to include the two-body part.  We assume this
additional interference because the two-body terms come from a scalar
operator with the same symmetry properties under parity and
time-reversal as $\Op_1$ and $\Op_3$, and because these terms are
independent of the WIMP spin $\spin_\chi$ (interference terms vanish
if they are linear in $\spin_\chi$~\cite{Fitzpatrick:2012ix}).
Therefore, Eq.~\eqref{structure_factors_def} neglects higher
multipoles $L=2$.  These are only non-vanishing for $^{131}$Xe (with a
$J=3/2$ ground state), but even in this case they are very small and
not coherent, as shown in Ref.~\cite{Vietze:2014vsa}.  An expression
similar to Eq.~\eqref{structure_factors_def}, only replacing the
nuclear response functions $\F_\pm$, $\F_\pi$, $\F_\pi^\theta$ associated with $L=0$
multipoles by the corresponding nuclear responses $\tilde{\F}_\pm$,
$\tilde{\F}_\pi$, $\tilde{\F}_\pi^\theta$ for $L=2$ can be added to the differential
WIMP--nucleus cross section above.

In order to justify Eq.~\eqref{structure_factors_def} we can consider
the more general differential cross section which accommodates the
contributions from all the NREFT operators that give rise to coherent
or quasi-coherent nuclear responses.  This involves the additional
operators
\beq
\Op_5=i\spin_\chi\cdot \big(\qq\times\vvp\big), \qquad \Op_8=\spin_\chi\cdot \vvp, \qquad \Op_{11}=i\spin_\chi\cdot \qq,
\label{subleading_op}
\eeq
which generate a nuclear $M$ response~\cite{Fitzpatrick:2012ix,Anand:2013yka}.
In this case, the generalized cross section reads
\begin{align}
\label{structure_factors_def_gen}
\frac{\diff \sigma_{\chi\N}^\text{SI}}{\diff \qq^2}&=\frac{\zeta^2}{4\pi\vv^2}\Bigg(
\: \bigg|\sum_{I=\pm}\Big[\xi_{\Op_1}f_I^{\Op_1}(\qq^2)\F_I^M(\qq^2)\notag\\
&+\xi_{\Op_3}f_I^{\Op_3}(\qq^2)\F_I^{\Phi''}(\qq^2)\Big]\notag\\
&+\xi_\pi f_\pi\F_\pi(\qq^2)+\xi_\pi^\theta f_\pi^\theta\F_\pi^\theta(\qq^2)\bigg|^2\notag\\
&+\sum_{i=5,8,11}\bigg|\sum_{I=\pm}\xi_{\Op_i}f_I^{\Op_i}\F_I^M(\qq^2)\bigg|^2 \:\Bigg).
\end{align}
The separation into kinematics $\xi_{\Op_i}$, nucleon form factors
$f_\pm^{\Op_i}$, and nuclear responses $\F(\qq^2)$ is chosen in
such a way that the form factors coincide with $f_N$ and $f_1^{V,N}$
as defined in Secs.~\ref{sec:scalar} and~\ref{sec:vector}.  The form
of the $\xi_{\Op_i}$, that set the scale for the $\Op_5$, $\Op_8$, and
$\Op_{11}$ contributions, originates in the NR expansion of the
effective operator to which they first contribute: the vector--vector,
axial-vector--vector, and pseudoscalar--scalar channels,
respectively~\cite{Hoferichter:2015ipa}
\begin{align}
\M_{1,\text{NR}}^{VV}(\Op_{5})&=f_1^{V,N}(t) \frac{\muN}{\mN}\frac{1}{\mc} \Op_5,\notag\\
\M_{1,\text{NR}}^{AV}(\Op_{8})&=2f_1^{V,N}(t) \Op_8, \notag\\
\M_{1,\text{NR}}^{PS}(\Op_{11})&= -f_N(t) \frac{1}{\mc}\Op_{11},
\end{align}
together with the operator multipole
decomposition~\cite{Fitzpatrick:2012ix,Anand:2013yka}. Altogether this
leads to
\begin{align}
\label{kin_factors}
\xi_{\Op_1}&=\xi_\pi=\xi_\pi^\theta=1,\qquad \xi_{\Op_3}=\frac{\qq^2}{2\mN^2},\\
\xi_{\Op_5}&=\frac{\muN|\qq||\vvp_T|}{2\mc\mN},\qquad \xi_{\Op_8}=|\vvp_T|,\qquad \xi_{\Op_{11}}=-\frac{|\qq|}{2\mc},\notag
\end{align}
with the corresponding form factors
\begin{align}
f_\pm^{\Op_1}(\qq^2)&=f_\pm^M(\qq^2),\qquad f_\pm^{\Op_3}(\qq^2)=f_\pm^{\Phi''},\notag\\
f_\pm^{\Op_5}&=f_\pm^{\Op_8}=\frac{1}{2}\Big[f_1^{V,p}\pm f_1^{V,n}\Big],\notag\\
f_\pm^{\Op_{11}}&=\frac{1}{2}\Big[f_p\pm f_n\Big],
\end{align}
where for the operators in Eq.~\eqref{subleading_op} only the leading
term has been listed.  The form factors for the $\Op_{5,8,11}$ terms
can be expressed in terms of the previously-defined quantities $f_N$
and $f_1^{V,N}$, since they first appear in the NR expansion of the
effective operators in Eq.~\eqref{Lagr} with scalar and vector nucleon
interactions, in a similar way as $\Op_1$ and $\Op_3$.

We note that Eq.~\eqref{structure_factors_def_gen} shows that the
$\Op_{5,8,11}$ operators do not interfere with $\Op_1$ or
$\Op_3$~\cite{Fitzpatrick:2012ix,Anand:2013yka}.  This is because
contrary to $\Op_1$ and $\Op_3$, the operators $\Op_{5}$, $\Op_{8}$,
and $\Op_{11}$ are linear in the WIMP spin $\spin_\chi$, and the
corresponding interference terms vanish after averaging over WIMP
spin-projections.  In addition, the kinematical factors imply that the
contributions of $\Op_{5,8,11}$ are suppressed by $|\vvp_T|$ or
$1/\mc$.  These two properties are crucial for the $\Op_3$ operator
being the main one-body correction to the standard SI analyses, as
anticipated in Eq.~\eqref{structure_factors_def}.  In the next
Sec.~\ref{sec:one-body} we show this explicitly by studying the
one-body structure factors for xenon isotopes.

\section{One-body currents}
\label{sec:one-body}

We calculate the structure factors as in our previous
work~\cite{Menendez:2012tm,Klos:2013rwa,Vietze:2014vsa}, by performing
large-scale shell-model calculations of all stable xenon isotopes in a
valence space comprising the $0g_{7/2}$, $1d_{5/2}$, $1d_{3/2}$,
$2s_{1/2}$, and $0h_{11/2}$ ($nlj$) orbitals for both neutrons and
protons, with $n$ the radial quantum number, $l$ the orbital angular
momentum in spectroscopic notation, and $j$ the total angular
momentum.  Our calculations therefore assume an isospin symmetric
$^{100}$Sn core.  For $^{132}$Xe, $^{134}$Xe, and $^{136}$Xe exact
diagonalizations are obtained in this valence space, while for the
remaining isotopes some truncations, which should not significantly
affect the nuclear ground states, are needed to keep the matrix
dimensions tractable, as discussed in
Refs.~\cite{Menendez:2012tm,Vietze:2014vsa}.  We use the shell-model
interaction GCN5082~\cite{Caurier:2007wq,Menendez:2008jp}, which has
also been used in neutrinoless double-beta decay calculations of
$^{136}$Xe~\cite{Caurier:2007wq,Menendez:2008jp}.  The low-energy
excitation spectra of all isotopes are very well
reproduced~\cite{Menendez:2012tm,Vietze:2014vsa}.  The
nuclear-structure calculations have been performed with the
shell-model code ANTOINE~\cite{Caurier:1999,Caurier:2004gf}.

The phenomenological nature of the shell-model interaction used makes
it difficult to estimate the theoretical uncertainties associated with
the nuclear-structure calculations.  Similarly, the systematic
uncertainty due to the truncations needed for some isotopes is
challenging to evaluate. It will be possible to address these aspects
with calculations based on ChEFT interactions, which provide natural
diagnostics to estimate nuclear-structure
uncertainties~\cite{Epelbaum:2014efa,Furnstahl:2015rha,Carlsson:2015vda,Simonis:2015vja}.
In the meantime, one measure for the reliability of the calculation
can be obtained by comparing the predicted excitation spectra with the
experimental results.

With the calculated xenon ground states we obtain all one-body nuclear
responses needed in Eq.~\eqref{structure_factors_def}.  The results,
summarized in Tables~\ref{tab:fits} and~\ref{tab:fits_L2}, are presented in
terms of the dimensionless parameter $u=\qq^2b^2/2$, where
$b=\sqrt{\hbar/\mN \omega}$ is the harmonic-oscillator length and
$\hbar\omega = (45 A^{-1/3}-25 A^{-2/3})\MeV$.  The nuclear response
functions leading to the structure factors are fit to the form
\beq
\label{fit_function}
\F(u) = e^{-\frac{u}{2}} \sum\limits_{i=0}^m c_i u^i,
\eeq
with $m=5$ for $\F^M_{\pm}$, $m=4$ for $\F^{\Phi''}_{\pm}$, and fixed
coefficients $c_0=A$ and $c_0=Z-N$ for $\F^M_{+}$ and $\F^M_{-}$,
respectively.  The form of the fit function follows the analytic
solution of the transition operators evaluated in the
harmonic-oscillator basis~\cite{Donnelly:1979ezn,Haxton:2008zza}.

The isoscalar nuclear $M$ operator has been known to lead to a
coherent contribution from all nucleons for a long time (at zero
momentum transfer)~\cite{Engel:1992bf}.  This justifies that the
nuclear response $\F^M_+$ associated with the $\Op_1$ operator is the
only one considered in most SI dark-matter direct-detection
analyses~\cite{Aprile:2012nq,Agnese:2014aze,Agnese:2015nto,Akerib:2015rjg,
Xiao:2015psa,Angloher:2015ewa,Amole:2016pye,Agnes:2015ftt,Armengaud:2016cvl}.

In turn, the nuclear $\Phi''$ operator, at zero momentum transfer, is
proportional to the sum over all nucleons of the single-nucleon
spin-orbit (${\bf l} \cdot {\bf s}$)
operator~\cite{Fitzpatrick:2012ix,Anand:2013yka}.  This implies that
nucleons in an orbital with spin parallel to the angular momentum,
$j=l+1/2$, contribute coherently.  Similarly, the nucleons in the
spin-orbit partner $j=l-1/2$ also contribute coherently, in such a way
that when both spin-orbit partners are filled their contributions
exactly cancel. However, in heavy nuclei the spin-orbit splitting is
important, with $j=l+1/2$ orbitals having significantly lower energies
than their spin-orbit partners.  In the case of xenon isotopes this
implies that the proton $0g_{9/2}$ and the neutron $0h_{11/2}$
orbitals are mostly filled (the latter especially for the more
neutron-rich isotopes), with the spin-orbit partners, proton
$0g_{7/2}$ and neutron $0h_{9/2}$ orbitals, mostly empty.  Therefore
the nuclear $\Phi''$ response, $\F^{\Phi''}$, shows a quasi-coherent
behavior~\cite{Fitzpatrick:2012ix,Anand:2013yka}, with the
contributions of about $20$ nucleons adding coherently in the
isoscalar case.  The total response is dominated by neutrons because
the $l=5$, $0h_{11/2}$ orbital accommodates $12$ nucleons, compared to
$10$ nucleons for the $l=4$, $0g_{9/2}$ orbital (the expectation value
of the single-particle spin-orbit operator is proportional to $l$ for
$j=l+1/2$ orbitals).  The nuclear response functions are larger for
the most neutron-rich isotopes with more neutrons in the $0h_{11/2}$
orbital.

The quasi-coherent nuclear response $\F^{\Phi''}$ is generated by the
$\Op_3$ operator.  In addition, in the total structure factor there is
an interference term between this contribution and the $\F^{M}$ term
from the dominant $\Op_1$ operator, as indicated by
Eq.~\eqref{structure_factors_def}~\cite{Fitzpatrick:2012ix,Anand:2013yka}.
This interference is important because, as discussed in
Sec.~\ref{sec:structure}, there is no other interference term coming
from one-body operators.  Altogether, the nuclear response
$\F^{\Phi''}$ generates the leading one-body-operator corrections to
the structure factors usually considered in SI analyses.

\begin{figure}[t] 
\centering
\includegraphics[width=\columnwidth,clip]{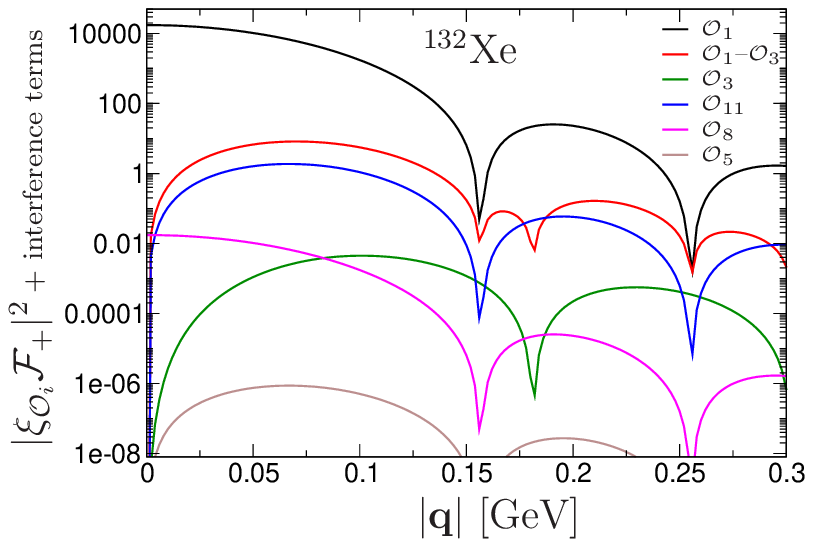}
\caption{Comparison of the isoscalar structure factors
associated with the coherent and quasi-coherent nuclear $M$ and $\Phi''$ 
responses.  The individual contributions corresponding to the $\Op_1$,
$\Op_3$, $\Op_{11}$, $\Op_{8}$, and $\Op_{5}$ operators, $|\xi_{\Op_i}
\F_\text{+}^{M/\Phi''}(\qq^2)|^2$, and the absolute value of the
$\Op_1$--$\Op_3$ interference term,
$|2\xi_{\Op_1}\xi_{\Op_3}\F^{M}_+(\qq^2)\F^{\Phi''}_+(\qq^2)|$, are
shown. For the evaluation of the structure factors associated with
$\Op_{11}$, $\Op_{8}$, and $\Op_{5}$ we take the relative WIMP
velocity $|\vvp_T|=10^{-3}$ and WIMP mass $\mc=2\GeV$, roughly the
minimal mass probed in xenon direct-detection experiments.  The
results, representative for all stable xenon isotopes, are shown for
the most abundant $^{132}$Xe.}
\label{fig:sf_isoscalar}
\end{figure}

\begin{figure}[t] 
\centering
\includegraphics[width=\columnwidth,clip]{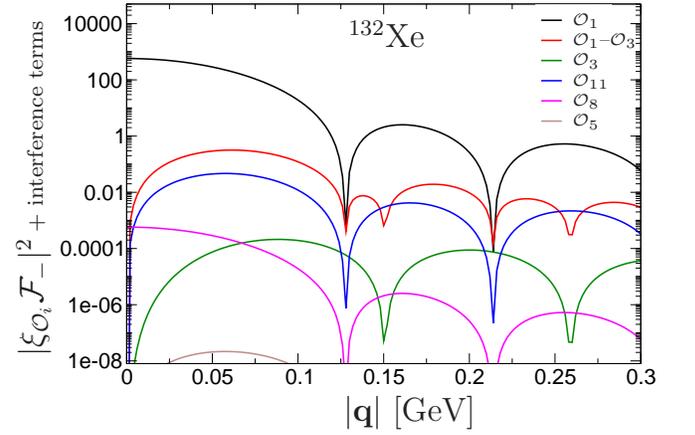}
\caption{Same as Fig.~\ref{fig:sf_isoscalar} but for the isovector case.
The isovector individual structure factors $|\xi_{\Op_i}
\F^{M/\Phi''}_-(\qq^2)|^2$, and the absolute value of the
$\Op_1$--$\Op_3$ interference term
$|2\xi_{\Op_1}\xi_{\Op_3}\F^{M}_-(\qq^2)\F^{\Phi''}_-(\qq^2)|$ are shown.}
\label{fig:sf_isovector}
\end{figure}

\begin{table*}[t]
\centering
\renewcommand{\arraystretch}{1.3}
\begin{tabular}{lccccccc}
\toprule
Isotope	& $^{128}$Xe & $^{129}$Xe & $^{130}$Xe & $^{131}$Xe  & $^{132}$Xe & $^{134}$Xe & $^{136}$Xe \\
\colrule
$J^{\Pi}$ & $0^+$ & $1/2^+$ & $0^+$ & $3/2^+$ & $0^+$ & $0^+$ & $0^+$ \\
\colrule
$b$~[fm] & $2.2847$ & $2.2873$ & $2.2899$ & $2.2925$ & $2.2950$ & $2.3001$ & $2.3051$ \\
\colrule
$c_1^{M+}$ & $ -126.455$ & $-128.09$ & $-129.753$ & $-131.26$ & $-132.835$ & $-135.861$ & $-138.787 $\\
$c_2^{M+}$ & $ 35.82$ & $36.4367$ & $37.2381$ & $37.8232$ & $38.4665$ & $39.6872$ & $40.9048 $\\
$c_3^{M+}$ & $ -3.66991$ & $-3.75317$ & $-3.89291$ & $-3.97171$ & $-4.06999$ & $-4.24713$ & $-4.41984 $\\
$c_4^{M+}$ & $ 0.125062$ & $0.129553$ & $0.139778$ & $0.142995$ & $0.149636$ & $0.159053$ & $0.165388 $\\
$c_5^{M+}$ & $ -5.63731 \times 10^{-4}$ & $-6.55816 \times 10^{-4}$ & $-9.30032 \times 10^{-4}$ & $-9.12955 \times 10^{-4}$ & $-0.00111463$ & $-0.00125724$ & $-0.00109211 $\\\colrule
$c_1^{M-}$ & $ 29.0588$ & $30.6854$ & $32.2019$ & $33.7021$ & $35.253$ & $38.2701$ & $41.2081 $\\
$c_2^{M-}$ & $ -11.7104$ & $-12.3687$ & $-13.1152$ & $-13.7433$ & $-14.4437$ & $-15.773$ & $-17.0848 $\\
$c_3^{M-}$ & $ 1.68447$ & $1.77928$ & $1.90775$ & $2.00031$ & $2.11305$ & $2.32061$ & $2.52635 $\\
$c_4^{M-}$ & $ -0.0820044$ & $-0.0868754$ & $-0.0948184$ & $-0.0991364$ & $-0.105689$ & $-0.116557$ & $-0.12686 $\\
$c_5^{M-}$ & $ 6.65781 \times 10^{-4}$ & $7.39474 \times 10^{-4}$ & $8.47975 \times 10^{-4}$ & $8.60686 \times 10^{-4}$ & $9.61344 \times 10^{-4}$ & $0.00106693$ & $0.00110965 $\\\colrule
$c_0^{\Phi''+}$ & $ -25.211$ & $-26.1264$ & $-27.7106$ & $-28.0443$ & $-28.7972$ & $-29.5095$ & $-29.8571 $\\
$c_1^{\Phi''+}$ & $ 17.592$ & $18.4401$ & $19.7108$ & $20.0888$ & $20.7751$ & $21.5578$ & $22.0402 $\\
$c_2^{\Phi''+}$ & $ -3.46466$ & $-3.64669$ & $-3.85805$ & $-3.94934$ & $-4.0995$ & $-4.27308$ & $-4.37033 $\\
$c_3^{\Phi''+}$ & $ 0.224722$ & $0.239379$ & $0.252667$ & $0.260624$ & $0.272865$ & $0.287393$ & $0.296134 $\\
$c_4^{\Phi''+}$ & $ -0.00353316$ & $-0.00399779$ & $-0.00444209$ & $-0.00468846$ & $-0.00507527$ & $-0.00555437$ & $-0.0059684 $\\\colrule
$c_0^{\Phi''-}$ & $ 3.89629$ & $5.47022$ & $6.28519$ & $6.90542$ & $7.93145$ & $9.3351$ & $10.1433 $\\
$c_1^{\Phi''-}$ & $ -4.73163$ & $-5.96963$ & $-6.63842$ & $-7.17962$ & $-8.01086$ & $-9.20279$ & $-9.96123 $\\
$c_2^{\Phi''-}$ & $ 1.48489$ & $1.7533$ & $1.85406$ & $1.97217$ & $2.12817$ & $2.35489$ & $2.48784 $\\
$c_3^{\Phi''-}$ & $ -0.140203$ & $-0.160094$ & $-0.166079$ & $-0.175248$ & $-0.186148$ & $-0.202364$ & $-0.212062 $\\
$c_4^{\Phi''-}$ & $ 0.00344765$ & $0.00387983$ & $0.00413453$ & $0.00437613$ & $0.00469887$ & $0.00519463$ & $0.00559688 $\\
\botrule
\end{tabular}
\renewcommand{\arraystretch}{1.0}
\caption{Spin/parity $J^{\Pi}$ of the nuclear ground states, 
harmonic-oscillator length $b$, and fit coefficients for the nuclear
response functions $\F_\pm^M$ and $\F_\pm^{\Phi''}$.  The fit
functions are $\F_\pm^{M}(u) = e^{-\frac{u}{2}}\sum_{i=0}^5 c_i^{M\pm} u^i$
(with $c_0=A$ and $c_0=Z-N$, respectively) and $\F_{\pm}^{\Phi''}(u) =
e^{-\frac{u}{2}} \sum_{i=0}^4 c_i^{\Phi''\pm} u^i$, with $u=\qq^2b^2/2$.  These
forms correspond to the analytical solution in the harmonic-oscillator
basis~\cite{Donnelly:1979ezn,Haxton:2008zza}.  For the $L=2$
multipoles in $^{131}$Xe, see Table~\ref{tab:fits_L2}.}
\label{tab:fits}
\end{table*}

\begin{table*}[t]
\centering
\renewcommand{\arraystretch}{1.3}
\begin{tabular}{lccccc}
\toprule
response & $\tilde c_1$ & $\tilde c_2$ & $\tilde c_3$ & $\tilde c_4$ & $\tilde c_5$\\\colrule
$\tilde{\F}_+^{M}$ & $2.17516$ & $-1.25386$ & $0.214567$ & $-0.0110964$ & $7.99074\times 10^{-5}$\\
$\tilde{\F}_-^{M}$ & $-0.344057$ & $0.208632$ & $-0.048112$ & $0.00351588$ & $-8.14509\times 10^{-5}$\\\colrule
& $\tilde c_0$ & $\tilde c_1$ & $\tilde c_2$ & $\tilde c_3$ & $\tilde c_4$\\\colrule
$\tilde{\F}_+^{\Phi''}$ & $0.498456$ & $-0.0289149$ & $-0.0160376$ & $-7.71842\times 10^{-5}$ & $4.59007\times 10^{-4}$\\
$\tilde{\F}_-^{\Phi''}$ & $-0.751871$ & $1.06826$ & $-0.227403$ & $0.00963627$ & $-4.14555\times 10^{-4}$\\
\botrule
\end{tabular}
\renewcommand{\arraystretch}{1.0}
\caption{Fit coefficients for the $L=2$ multipoles in $^{131}$Xe, 
parameterized by $\tilde{\F}_\pm^{M}(u) = e^{-\frac{u}{2}}\sum_{i=1}^5
\tilde c_i u^i$, $\tilde{\F}_{\pm}^{\Phi''}(u) = e^{-\frac{u}{2}}
\sum_{i=0}^4 \tilde c_i u^i$. Notation and oscillator length as in
Table~\ref{tab:fits}.}
\label{tab:fits_L2}
\end{table*}

This is illustrated in Figs.~\ref{fig:sf_isoscalar}
and~\ref{fig:sf_isovector}, which compare for the isoscalar and
isovector cases, respectively, the structure factors associated with
the coherent and quasi-coherent nuclear $M$ and $\Phi''$ responses
generated by the operators $\Op_1$, $\Op_3$, $\Op_{11}$, $\Op_{8}$,
and $\Op_{5}$.  In this comparison the values of the associated
nucleon couplings and form factors are not included, so some caution
needs to be taken in the interpretation of the figures due to
differences in the combination of the Wilson coefficients for the
different contributions. However, the results are shown on a
logarithmic scale, and the main features in
Figs.~\ref{fig:sf_isoscalar} and~\ref{fig:sf_isovector} should still
be valid once all corresponding couplings and form factors are
included.

Figure~\ref{fig:sf_isoscalar} shows that the standard SI structure
factor proportional to $A^2$, originating from the $\Op_1$ operator,
receives the leading one-body correction from the interference with
the $\F^{\Phi''}$ response due to $\Op_3$.  This correction is only of
the order of $1$ per mil because $\F^{\Phi''}$ comes with a
kinematical factor $\xi_{\Op_3}=\qq^2/2\mN^2$.  Consequently, the
interference term vanishes at $|\qq|=0$.

The next contribution in this hierarchy comes from the nuclear $M$
response originating from the $\Op_{11}$ operator.  Due to the
associated kinematical factor $\xi_{\Op_{11}}=|\qq|/\mc$, this
contribution also vanishes at $|\qq|=0$, and becomes less important
for heavier WIMPs.  Figure~\ref{fig:sf_isoscalar} shows the results
for $\mc=2\GeV$, roughly the smallest WIMP mass probed by xenon
direct-detection experiments.  For heavier WIMPs the structure factor
associated with the $\Op_{11}$ operator is reduced, and for
$\mc\approx 50\GeV$ this structure factor is comparable to the one
corresponding to the $\Op_3$ operator, originating solely from the
nuclear $\F^{\Phi''}$ response.  The latter structure factor is
suppressed by three additional orders of magnitude compared to the
leading correction to the standard SI structure factor, the
$\Op_1$--$\Op_3$ interference term.

Finally, the structure factors coming from the nuclear $M$ responses
associated with the $\Op_8$ and $\Op_5$ operators are even smaller,
because they are suppressed by the very small WIMP velocity
$|\vvp_T|\approx10^{-3}$ in their kinematical factors, see
Eq.~\eqref{kin_factors}.  Note that, as emphasized in
Refs.~\cite{Fitzpatrick:2012ix,Anand:2013yka}, the $\Op_3$ operator,
similarly to $\Op_8$ and $\Op_5$, involves the velocity operator
$\vvp$, but for $\Op_3$ the associated nuclear operator does not
depend on the WIMP velocity with respect to the center-of-mass,
$\vvp_T$, but on the nucleon's velocity operator, which is part of the
$\Phi''$ operator and generates a milder suppression factor
$|\qq|/\mN$.

The isovector results shown in Fig.~\ref{fig:sf_isovector} are very
similar to the isoscalar case.  The only difference is that all
structure factors are smaller because in this case the contributions
of protons and neutrons partially cancel.

Similarly to this generalized SI analysis, the standard SD structure
factor will receive additional contributions beyond the $\Op_4$ and
$\Op_6$ operators. In particular, the $\Op_3$, $\Op_7$, $\Op_9$, and
$\Op_{10}$ operators contribute to $\Sigma'$ or $\Sigma''$, and
$\Op_5$, $\Op_8$ to the additional $\Delta$ response. In addition
there will be $\Op_4$--$\Op_5$ and $\Op_8$--$\Op_9$
$\Sigma'$--$\Delta$ interference terms~\cite{Fitzpatrick:2012ix}. All
these additional contributions vanish at $|\qq|=0$, except for the
$\Op_7$ response which is suppressed by the WIMP velocity $|\vvp_T|
\approx 10^{-3}$. 
Note also that only $\Op_5$ interferes
with the dominant SD response, but this operator only
appears at higher (fourth) order in ChEFT~\cite{Hoferichter:2015ipa}.
Likewise, the $\tilde\Phi'$ response receives contributions from higher
ChEFT orders only. Therefore these corrections to SD scattering are
expected to be small. We defer a detailed analysis of generalized SD
scattering to future work.

\section{Two-body currents}
\label{sec:two-body}

As discussed in Sec.~\ref{sec:one-body} the shell-model calculations
are based on a core, while the many-body problem is explicitly solved
for nucleons close to the Fermi level in the valence space. This
generally leads to very good agreement to experiment for
spectroscopy~\cite{Caurier:2004gf}, including the isotopes relevant
for dark-matter direct detection~\cite{Klos:2013rwa,Vietze:2014vsa}.

However, for the standard SI scattering (nuclear $M$ response) all
nucleons contribute coherently, so that the bulk of the nuclear
response is in fact generated by the inert core.  A similar argument
can be made for the quasi-coherent $\Phi''$ response in xenon, where
the core protons in the $0g_{9/2}$ orbital are responsible for about
half of the total response.  The relatively small sensitivity of these
nuclear responses to the nuclear structure was discussed in
Ref.~\cite{Vietze:2014vsa}, and justifies the use of the simple Helm
form factor [see Eq.~\eqref{Helm}] in the standard SI analysis.

\begin{figure}[t] 
\centering
\includegraphics[width=\columnwidth,clip]{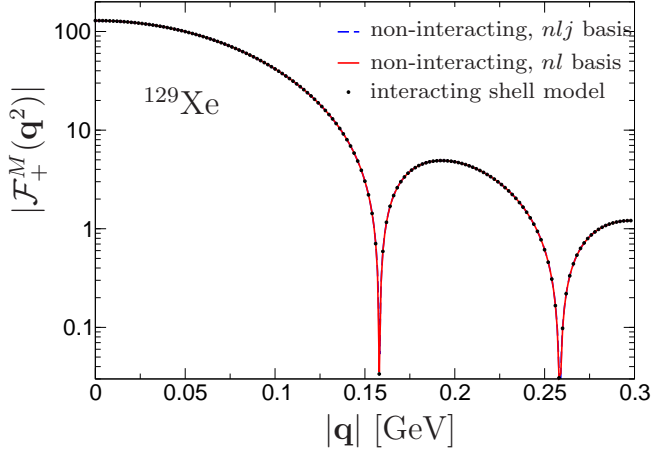}
\caption{$\F^M_{+}(\qq^2)$ for $^{129}$Xe obtained from three different 
approximations: shell-model calculation from~\cite{Vietze:2014vsa} (black 
dots), non-interacting shell model with $j$-coupling (red solid line), and 
in $nl$ basis (blue dashed line).}
\label{fig:naive_SM_Xe129}
\end{figure}

In addition, the nuclear response can be calculated in a
non-interacting shell-model picture, where only the lowest-lying
orbitals are filled with particles.  Figure~\ref{fig:naive_SM_Xe129}
shows the $\F_+^M(\qq^2)$ response for $^{129}$Xe, using a
non-interacting shell model and single-particle orbitals with and
without $j$-coupling (but with occupation numbers from the interacting
shell model, see Table~\ref{tab:xenon129}).  The agreement with the
full shell-model calculation is very good, showing that the dependence
on correlations among the valence nucleons as well as $j$-coupling
effects are small for this response. Likewise, the effect of using
naive or shell-model-based occupation numbers would be hardly visible
in the figure.

\begin{table}[t]
\centering
\renewcommand{\arraystretch}{1.3}
\begin{tabular}{ccccc}
\toprule
$n$ & $l$ & maximum occupation & $n_{nl}^p$ & $n_{nl}^n$\\\colrule
$0$ & $0$ & $2$ &  $1$ & $1$\\ 
$0$ & $1$ & $6$ &  $1$ & $1$\\ 
$0$ & $2$ & $10$ & $1$ & $1$\\ 
$1$ & $0$ & $2$ &  $1$ & $1$\\ 
$0$ & $3$ & $14$ & $1$ & $1$\\ 
$1$ & $1$ & $6$  & $1$ & $1$\\ 
$0$ & $4$ & $18$ & $0.68$ & $0.99$\\ 
$1$ & $2$ & $10$ & $0.16$ & $0.79$\\ 
$2$ & $0$ & $2$ &  $0.06$ & $0.58$\\ 
$0$ & $5$ & $22$ & $0.01$ & $0.37$\\ 
\botrule
\end{tabular}
\renewcommand{\arraystretch}{1.0}
\caption{Relative occupation numbers $n^\tau_{nl}$ for the $nl$ orbitals in
$^{129}$Xe and maximum occupation including spin degeneracy.  For
orbitals in the valence space, the results of the shell-model
diagonalization are used.}
\label{tab:xenon129}
\end{table}

In view of these findings, we evaluate the two-body matrix elements of
Eq.~\eqref{scalar_two_body} by
\begin{align}
\label{M2N_start}
\F_\pi(\qq^2)&=\frac{1}{2}\sum_{\text{occ}}\langle N_1N_2|(1-P_{12})|\frac{1}{f_\pi}\M_{2,\text{NR}}^{SS}|N_1N_2\rangle,\notag\\
|N_1N_2\rangle&=|n_1l_1m_1\sigma_1\tau_1n_2l_2m_2\sigma_2\tau_2\rangle,
\end{align}
and analogously for $\F_\pi^\theta(\qq^2)$,
where the sum runs over occupied states (e.g., for $^{129}$Xe according to
Table~\ref{tab:xenon129}) and $P_{12} = P_k P_\sigma P_\tau$ is the
exchange operator with
\beq
P_\sigma=\frac{1}{2}\big(\unity+\sig_1\cdot\sig_2\big),\qquad P_\tau = \frac{1}{2}\big(\unity+\ttau_1\cdot\ttau_2\big),
\eeq
and $P_k$ exchanges the momenta. Summing over spins $\sigma_i$ and 
evaluating the matrix element in Eq.~\eqref{M2N_start}
in the harmonic-oscillator basis, we obtain
\begin{align}
\label{naive_SM_Fpi}
\F&_\pi(\qq^2)=\frac{\mpi}{2}\bigg(\frac{g_A}{2F_\pi}\bigg)^2
\sum_{n_1l_1n_2l_2}\sum_{\tau_1 \tau_2}\int\frac{\diff^3 p_1\diff^3 p_2\diff^3 p_1'\diff^3 p_2'}{(2\pi)^6}\notag\\
&\times R_{n_1l_1}(|\pp_1'|)R_{n_2l_2}(|\pp_2'|)R_{n_1l_1}(|\pp_1|)R_{n_2l_2}(|\pp_2|)\notag\\
&\times\frac{(2l_1+1)(2l_2+1)}{16\pi^2}P_{l_1}\big(\hat \pp_1'\cdot \hat \pp_1\big)P_{l_2}\big(\hat \pp_2'\cdot \hat \pp_2\big)\notag\\
&\times (2\pi)^3\delta^{(3)}\big(\pp_1+\pp_2-\pp_1'-\pp_2'-\qq\big)\notag\\
&\times\big(3-\ttau_1\cdot\ttau_2\big)
\frac{\qq_1^\text{ex}\cdot \qq_2^\text{ex}}{\big((\qq_1^\text{ex})^2+\mpi^2\big)\big((\qq_2^\text{ex})^2+\mpi^2\big)},
\end{align}
with
\beq
\qq_1^\text{ex}=\pp_2'-\pp_1,\quad \qq_2^\text{ex}=\pp_1'-\pp_2,\quad \qq=-\qq_1^\text{ex}-\qq_2^\text{ex}, 
\eeq
and radial wave functions
\beq
R_{nl}(k)=b^{3/2}\sqrt{\frac{2\,n!}{\Gamma(n+l+3/2)}}(bk)^le^{-\frac{(bk)^2}{2}}L_n^{l+1/2}\big[(bk)^2\big].
\eeq
The expression for $\F_\pi^\theta(\qq^2)$ is analogous.
The sum over $m_1$, $m_2$ has been performed using the addition theorem for the spherical harmonics,
assuming an equal filling of all orbitals with different $m$ projections.
Apart from the momentum integrals, which can be performed numerically for given $\{n_1l_1n_2l_2\}$, only the isospin part of Eq.~\eqref{naive_SM_Fpi} needs to be evaluated. This leads to
\begin{align}
&\sum_{n_1l_1n_2l_2}\sum_{\tau_1 \tau_2} \big(3-\ttau_1\cdot\ttau_2\big) \\
&=2\sum_{n_1l_1n_2l_2}\big[n_{n_1l_1}^p n_{n_2l_2}^p+n_{n_1l_1}^n n_{n_2l_2}^n+4n_{n_1l_1}^p n_{n_2l_2}^n\big], \notag
\end{align}
where the $n^\tau_{nl}$ denote the relative occupation numbers of a given orbital.
In Table~\ref{tab:xenon129} we list these occupation numbers for the
case of $^{129}$Xe used in the calculation of $\F_\pi$ and $\F_\pi^\theta$ as well as the
$nl$-basis calculation shown in Fig.~\ref{fig:naive_SM_Xe129}.  For
orbitals in the valence space of the shell-model calculations, the
result of the full diagonalization is used, even though the
sensitivity to this is minor.

The results for $\F_\pi$ and $\F_\pi^\theta$ can be fit with the same functional form as
given in Eq.~\eqref{fit_function} for the one-body case, see
Table~\ref{tab:fits_2b} for the corresponding coefficients.  Keeping
terms up to $m=5$ provides the best description also for the two-body
terms.  This form can be expected based on normal-ordering arguments:
after the summation over the second nucleon, the result only depends
on $\pp_1$, $\pp_1'$, and $\sig_1$, so that the corresponding
operators depend on $\qq$, $\vvp$, and $\spin_N$, and can be written
in terms of $\Op_1$, $\Op_3$, as well as other operators subleading in
our analysis.  Then the summation over spins performed before
Eq.~\eqref{naive_SM_Fpi} eliminates the dependence on $\Op_3$ (as well
as higher multipoles).  Therefore, apart from suppressed
contributions, we expect the normal ordering to reduce the two-body
matrix element to a one-body matrix element of $\Op_1$, with
corresponding fit function as given in Eq.~\eqref{fit_function} with $m=5$.

\begin{table*}[t]
\centering
\renewcommand{\arraystretch}{1.3}
\begin{tabular}{lccccccc}
\toprule
Isotope	& $^{128}$Xe & $^{129}$Xe & $^{130}$Xe & $^{131}$Xe  & $^{132}$Xe & $^{134}$Xe & $^{136}$Xe \\
\colrule
$J^{\Pi}$ & $0^+$ & $1/2^+$ & $0^+$ & $3/2^+$ & $0^+$ & $0^+$ & $0^+$ \\
\colrule
$b$~[fm] & $2.2847$ & $2.2873$ & $2.2899$ & $2.2925$ & $2.2950$ & $2.3001$ & $2.3051$ \\
\colrule
$c_0^\pi$ & $ -2.42605$ & $-2.44233$ & $-2.45715$ & $-2.47546$ & $-2.49308$ & $-2.52965$ & $-2.56752 $\\
$c_1^\pi$ & $ 2.01883$ & $2.03693$ & $2.063$ & $2.08643$ & $2.11087$ & $2.15556$ & $2.19645 $\\
$c_2^\pi$ & $ -0.576294$ & $-0.579809$ & $-0.594377$ & $-0.602812$ & $-0.612728$ & $-0.62789$ & $-0.642445 $\\
$c_3^\pi$ & $ 0.077613$ & $0.0775201$ & $0.0810307$ & $0.0824072$ & $0.0844652$ & $0.0863288$ & $0.0883411 $\\
$c_4^\pi$ & $ -0.00519097$ & $-0.00512894$ & $-0.0055788$ & $-0.00570646$ & $-0.00597987$ & $-0.00602651$ & $-0.00611004 $\\
$c_5^\pi$ & $ 1.39081 \times 10^{-4}$ & $1.35327 \times 10^{-4}$ & $1.59249 \times 10^{-4}$ & $1.65335 \times 10^{-4}$ & $1.82198 \times 10^{-4}$ & $1.78002 \times 10^{-4}$ & $1.75076 \times 10^{-4} $\\
\colrule
$c_0^\theta$ & $ -24.8768$ & $-25.039$ & $-25.2034$ & $-25.3895$ & $-25.5691$ & $-25.9446$ & $-26.3396 $\\
$c_1^\theta$ & $ 18.5427$ & $18.8087$ & $18.9813$ & $19.2032$ & $19.4359$ & $19.8659$ & $20.248 $\\
$c_2^\theta$ & $ -4.81514$ & $-4.90161$ & $-4.96798$ & $-5.03573$ & $-5.11592$ & $-5.2492$ & $-5.38323 $\\
$c_3^\theta$ & $ 0.631787$ & $0.644029$ & $0.650297$ & $0.658108$ & $0.670645$ & $0.683946$ & $0.70754 $\\
$c_4^\theta$ & $ -0.0477761$ & $-0.0488906$ & $-0.0483377$ & $-0.0487362$ & $-0.0500243$ & $-0.049659$ & $-0.0522969 $\\
$c_5^\theta$ & $ 0.00171469$ & $0.0017729$ & $0.00165885$ & $0.00167317$ & $0.00174703$ & $0.00163541$ & $0.001781 $\\
\botrule
\end{tabular}
\renewcommand{\arraystretch}{1.0}
\caption{Spin/parity $J^{\Pi}$ of the nuclear ground states, 
harmonic-oscillator length $b$, and fit coefficients for the nuclear
response functions $\F_\pi$ and $\F_\pi^\theta$, with fit functions $\F_\pi(u) =
e^{-\frac{u}{2}}\sum_{i=0}^5 c_i^\pi u^i$, $\F_\pi^\theta(u) =
e^{-\frac{u}{2}}\sum_{i=0}^5 c_i^\theta u^i$, and $u=\qq^2b^2/2$.}
\label{tab:fits_2b}
\end{table*}

We note that the equal-filling approximation picks out the $L=0$ part
of the response, as required for the decomposition of the SI structure
factor given in Eq.~\eqref{structure_factors_def}. 
The $L=2$ multipole contribution, only relevant for $^{131}$Xe, would
only appear as a correction to the strongly suppressed one-body $L=2$
structure factor, which itself enters below the
$\Op_{11}$ curve in Fig.~\ref{fig:sf_isoscalar}. Therefore it can be safely
neglected. 

The $\F_\pi(0)$ contribution has been considered before in
Refs.~\cite{Prezeau:2003sv,Cirigliano:2012pq,Cirigliano:2013zta},
based on results for closed-shell nuclei and represented in terms of a
fit linear in $A$.  In our conventions, the results for $A=132$ are
$\F_\pi(0)=-2.4(0.8)$~\cite{Prezeau:2003sv},
$\F_\pi(0)=-1.4$~\cite{Cirigliano:2012pq}, and
$\F_\pi(0)=-1.9$~\cite{Cirigliano:2013zta}, in reasonable agreement
with our value.  The remaining differences can be traced back to our
improved nuclear structure calculation and additional corrections from
modeling nuclear short-range correlations~\cite{Vincenzo:private}
included in
Refs.~\cite{Prezeau:2003sv,Cirigliano:2012pq,Cirigliano:2013zta}. The
latter are not dictated by ChEFT in this form, and thus not present in
our calculation.  This strategy is in agreement with findings for
nuclear matrix elements of neutrinoless double-beta
decay~\cite{Menendez:2011qq,Engel:2014pha}, where the effects of
short-range correlations are small after the momentum dependence of
the one-body currents is included.

\begin{figure}[t] 
\centering
\includegraphics[width=\columnwidth,clip]{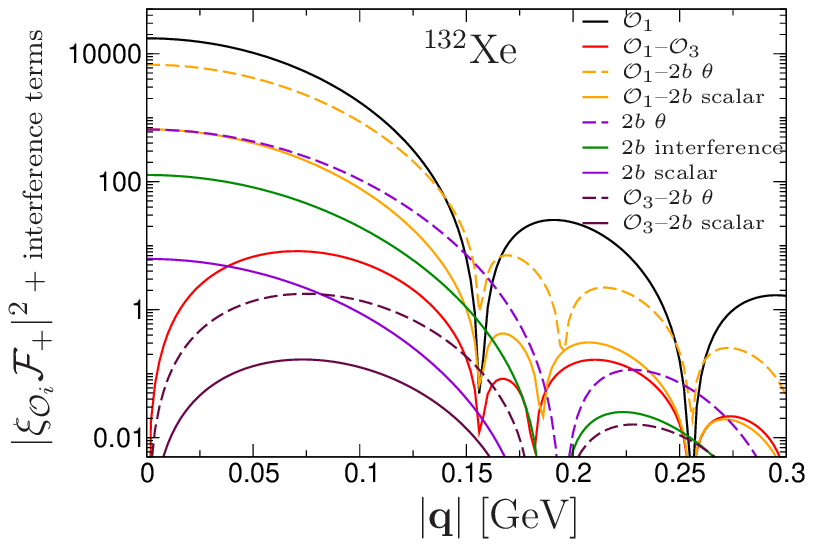}
\caption{Same as Fig.~\ref{fig:sf_isoscalar}, but including the 
two-body-current contribution $|\xi_\pi^{(\theta)}\F_\pi^{(\theta)}(\qq^2)|^2$ 
as well as the interference terms $|2\xi_{\Op_1}\xi_\pi^{(\theta)}\F^{M}_+(\qq^2)
\F_\pi^{(\theta)}(\qq^2)|$ and $|2\xi_{\Op_3}\xi_\pi^{(\theta)}\F^{\Phi''}_+(\qq^2)\F_\pi^{(\theta)}(\qq^2)|$. 
Solid (dashed) lines refer to $\F_\pi$ ($\F_\pi^\theta$). The green line indicates the interference
$|2\xi_\pi\xi_\pi^{\theta}\F_\pi(\qq^2)\F_\pi^{\theta}(\qq^2)|$ of the two-body terms.
The responses associated with $\Op_3$, $\Op_5$, $\Op_8$, and $\Op_{11}$ have been 
omitted for clarity.}
\label{fig:sf_isoscalar_2B}
\end{figure}

The consequences for the structure factors are illustrated in
Fig.~\ref{fig:sf_isoscalar_2B}, an extension of
Fig.~\ref{fig:sf_isoscalar} that includes the effect of
$\F_\pi(\qq^2)$ and $\F_\pi^\theta(\qq^2)$ as well as the interference terms with the isoscalar
one-body operators. Figure~\ref{fig:sf_isoscalar_2B} shows that the
two-body contributions constitute the leading correction to the $\Op_1$
structure factor. In particular, $\F_\pi^\theta(0)$ surpasses $\F_\pi(0)$
by an order of magnitude, to end up at a similar level as the isovector one-body contribution.
Equation~\eqref{scalar_two_body} illustrates the reason for this enhancement:
the factor $4$ from the momentum-independent term and the fact that 
the integral over $-\qq_1\cdot \qq_2/\mpi^2$ adds an additional factor about $3$
combine to the final factor of $10$.
It is also important to note that, in contrast to
the structure factor associated with $\Op_3$ (including its
interference with $\Op_1$) the two-body structure factors do not
vanish at $|\qq|=0$.

Even though the main hierarchy suggested by Fig.~\ref{fig:sf_isoscalar_2B}
should be relatively general, we stress that this comparison assumes that the
nucleon form factors are all of roughly the same size,
and that additional relative suppressions and enhancements may occur,
as for instance indicated by the simple models explored in Sec.~\ref{sec:scalar_ud},
where the relative size of both two-body terms is seen to be similar
due to the large single-nucleon matrix element
that compensates for the larger $\F_\pi^\theta(0)$.

Also, when comparing the hierarchy of isoscalar and isovector responses,
one should keep in mind that for theories with an approximate isospin
symmetry, there could be an additional suppression hidden, e.g., in
$f_p-f_n$. In either case, the dominant contribution will actually be
generated by the interference term $|2\F^{M}_+(\qq^2)\F^{M}_-(\qq^2)|$
with the isoscalar response.  In addition to the hierarchies studied
in Figs.~\ref{fig:sf_isoscalar}, \ref{fig:sf_isovector}, and
\ref{fig:sf_isoscalar_2B}, there are also $\qq^2$-dependent
corrections to the one-body form factors, which we address in the
following section.

\section{Parameters in general spin-independent scattering}
\label{sec:parameters_radius}

The hierarchy of the one- and two-body contributions discussed in
Secs.~\ref{sec:one-body} and~\ref{sec:two-body}, combined with the
general expression for the structure factor in
Eq.~\eqref{structure_factors_def} determines the number of independent
parameters in our analysis of general SI scattering.  First, however,
we need to quantify the momentum-dependent corrections to the one-body
form factors reviewed in Secs.~\ref{sec:scalar} and~\ref{sec:vector},
including the scalar radii, anomalous magnetic moments, as well as
strangeness radii and moments.  We refer generically to all these
contributions as radius corrections. They are evaluated in the
following Sec.~\ref{sec:radius}.  In Sec.~\ref{sec:parameters} we then
discuss the number of independent parameters appearing in the analysis
of general SI scattering, and as examples, in
Sec.~\ref{sec:scalar_ud} we focus on two simple cases:
the case of scalar interactions with $u$ and $d$ quarks only,
and purely gluonic couplings.

\subsection{Radius corrections}
\label{sec:radius}

The chiral counting that underlies the decomposition in
Eq.~\eqref{structure_factors_def} and Eq.~\eqref{form_factors} implies
that radius corrections are expected to contribute at a similar level
as the leading two-body currents. Moreover, since only the
coefficients of $\F_\pm^M(\qq^2)$ are affected, these corrections
concern the response of the $\Op_1$ operator, being coherently
enhanced. By definition, radius corrections vanish for vanishing
momentum transfer, but they could become relevant for larger $\qq^2$
values. The exact shape depends on the underlying BSM physics as well
as on their relative size compared to the leading nucleon form factor,
e.g., as seen in Eq.~\eqref{vector_couplings}, in the case of
$C_s^{VV}$ the leading contribution vanishes and radius corrections
generate all sensitivity to this Wilson coefficient.

\begin{figure}[t] 
\centering
\includegraphics[width=\columnwidth,clip]{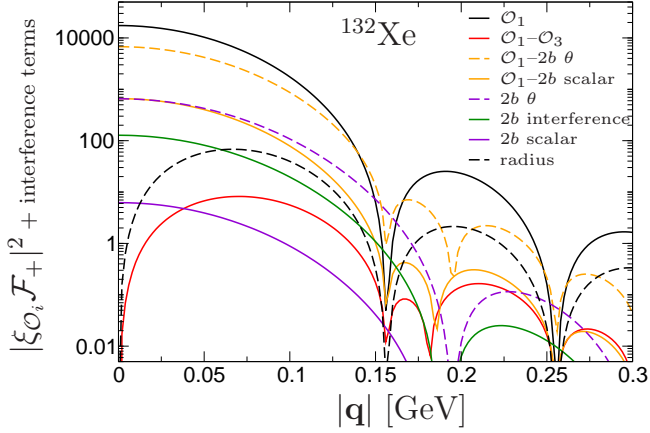}
\caption{Same as Fig.~\ref{fig:sf_isoscalar_2B}, but including the 
generic size of radius corrections (black dashed line), as discussed
in the text. Note that the $\Op_3$--$2b$ interference terms have been dropped.}
\label{fig:sf_isoscalar_2B_radius}
\end{figure}

In order to estimate the generic size of radius corrections in a
simple way, we factor out the nucleon mass as a representative
hadronic scale, leading to a typical $\qq^2/\mN^2$ suppression in the
associated structure factor.  This is illustrated in
Fig.~\ref{fig:sf_isoscalar_2B_radius} by means of the interference
term of radius corrections with $\Op_1$ (again assuming that the
remaining coefficients are both equal to $1$).  As expected, the
correction is irrelevant at $|\qq|=0$, but it is one of the largest
contributions for finite $|\qq|$, only second to the $\Op_1$--two-body
interference and $|\F_\pi^\theta|^2$ (and thus also below the interference with the isovector
$\Op_1$ operator not shown in Fig.~\ref{fig:sf_isoscalar_2B_radius}).
In particular Fig.~\ref{fig:sf_isoscalar_2B_radius} shows that the
radius corrections are expected to be more important than the
interference of the standard SI response with the new NREFT operator
$\Op_3$. This estimate supports the expectation from ChEFT that radius
corrections need to be included on the same footing as higher-order
momentum-dependent operators.

\subsection{Independent parameters}
\label{sec:parameters}

Within the formalism put forward in Sec.~\ref{sec:formalism}, the
decomposition of the WIMP--nucleus cross
section in Eq.~\eqref{structure_factors_def} therefore involves $8$
parameters that can be extracted from the dependence on
$Z$, $N$, and $|\qq|$, i.e., from direct-detection measurement on
different nuclear targets.  These are
\begin{enumerate}
\item two (isoscalar and isovector) leading coefficients of the $M$ response
\beq
\label{cMpm}
c_\pm^M=\frac{\zeta}{2}\Big[f_p\pm f_n+f_1^{V,p}\pm f_1^{V,n}\Big],
\eeq
\item two coefficients of the two-body responses
\beq
\label{cpi}
c_\pi=\zeta f_\pi,\qquad c_\pi^\theta=\zeta f_\pi^\theta,
\eeq
\item two (isoscalar and isovector) radius corrections to the $M$ response
\beq
\label{dotcMpm}
\dot c_\pm^M=\frac{\zeta m_N^2}{2}\bigg[\dot f_p\pm \dot f_n+\dot f_1^{V,p}\pm \dot f_1^{V,n}
+\frac{1}{4\mN^2}\Big(f_2^{V,p}\pm f_2^{V,n}\Big)\bigg],
\eeq
\item two (isoscalar and isovector) coefficients of the $\Phi''$ response
\beq
\label{cPhipm}
c_\pm^{\Phi''}=\frac{\zeta}{2}\Big(f_2^{V,p}\pm f_2^{V,n}\Big).
\eeq
\end{enumerate}
These $8$ parameters are not all independent,
since they map onto the seven Wilson coefficients $C^{SS}_q$,
$C'^{S}_g$, and $C^{VV}_q$ (with $q=u,d,s$) for a Dirac WIMP,
which reduces to four in the Majorana case where the $C^{VV}_q$ vanish.
Indeed, if higher orders in the momentum expansion
or $\eta$-exchange currents were considered, the number of parameters
in the decomposition of the nucleon form factors and two-body currents
would be even larger, so that in general a correlated analysis is called for.

The discussion of hierarchies in terms of
Figs.~\ref{fig:sf_isoscalar}, \ref{fig:sf_isovector},
\ref{fig:sf_isoscalar_2B}, and \ref{fig:sf_isoscalar_2B_radius} also
shows that if the minimal extension of the standard SI response is
sought, the analysis should include $c_\pm^M$, $c_\pi$, and $c_\pi^\theta$, extending
the standard formalism by the leading isovector and two-body
responses. These findings provide the basis for the discussion of
the general SI analysis strategy for direct-detection experiments
formulated in Sec.~\ref{sec:overview}.

\subsection{Examples: scalar interactions with $\boldsymbol{u}$ and $\boldsymbol{d}$ quarks and purely gluonic couplings}
\label{sec:scalar_ud}

Further simplifications can occur if specific assumptions are made
about the Wilson coefficients. As an example, we first consider the case of
purely scalar interactions, $C_q^{VV}=0$, with $u$ and $d$ quarks
only, i.e., with $C_s^{SS}=C'^S_{g}=0$. In this case, the
non-vanishing hadronic coefficients in Eq.~\eqref{form_factors}
become related according to
\begin{align}
\label{scalar_ud}
\frac{f_p+f_n}{2}&=\frac{\sigma_{\pi N}}{\mpi}f_\pi= 0.43(3)f_\pi,\notag\\
\frac{f_p-f_n}{2}&=-\frac{2Bc_5(m_d-m_u)}{\xi\mpi}\tilde f_\pi= 0.020(5) \tilde f_\pi,\notag\\
\frac{\dot f_p+\dot f_n}{2}&=\frac{\dot\sigma}{\mpi}f_\pi= 1.72(6) \mN^{-2}f_\pi.
\end{align}
Therefore, there are only two linearly independent parameters, namely
the isoscalar and the isovector coupling to the nucleon ($f_p \pm f_n$
or equivalently $f_\pi$ and $\tilde f_\pi$); the other
parameters, the coupling to the pion and the nucleon radius
corrections are then fully determined. To calculate the coefficients in
the above equations, we have used $\sigma_{\pi N}=59.1(3.5)\MeV$
from Ref.~\cite{Hoferichter:2015dsa}, $\dot\sigma$ and $\xi$ as given in
Sec.~\ref{sec:scalar}, and $B c_5(m_d-m_u)=-0.51(8)\MeV$ as extracted
from the electromagnetic proton--neutron mass difference
$(m_p-m_n)^\text{em}=0.76(30)\MeV$ via the Cottingham
formula~\cite{Cottingham:1963zz,Gasser:1974wd,Gasser:2015dwa} (and
consistent with lattice determinations~\cite{Borsanyi:2014jba}). The
new hadronic coefficient (in addition to the standard SI analysis) is
then given by
\beq
\tilde f_\pi=\frac{\mpi}{\Lambda^3}\Big(C^{SS}_{u}f_u^\pi-C^{SS}_{d}f_d^\pi\Big),
\eeq
thus differing by the relative sign from $f_\pi$ [see
Eq.~\eqref{fpi}].  According to Eq.~\eqref{scalar_ud}, the isoscalar
hadronic form factor and its radius correction are of the same
size as the two-body coefficient $f_\pi$ up to a factor of $2$, while the
isovector contribution is further suppressed by an order of magnitude
(the size of such isospin-violating effects has been studied in the
context of simplified models in Ref.~\cite{Crivellin:2015bva}), unless
this suppression in the hadronic input is balanced by $\tilde f_\pi
/f_\pi$.

At $|\qq|=0$, the dominant correction to the standard SI response is
thus generated by the two-body current, a reduction of the
WIMP--nucleus cross section by about\footnote{The very large effects 
of up to $60\%$ quoted in Ref.~\cite{Prezeau:2003sv} rely on a
specific parameter choice $r \approx 1$. For the example considered
here, this implies $2f_\pi/(f_p+f_n) \approx \mpi/(m_u+m_d) \approx
17$, in contradiction to Eq.~\eqref{scalar_ud}. For realistic values
of the hadronic couplings the two-body corrections are of the expected
size of $(5\text{--}10)\%$, while enhancements are possible if cancellations
in the leading contribution occur.}
\beq
\label{corr_2b}
2\frac{2f_\pi}{f_p+f_n}\frac{\F_{\pi}(0)}{A}=-9\%,
\eeq
followed by the isovector contribution, which affects the rate by 
\beq
\label{corr_IV}
2\frac{f_p-f_n}{f_p+f_n}\frac{Z-N}{A}=-2\% \frac{\tilde f_\pi}{f_\pi}.
\eeq

As a second example, it is also instructive to consider the case of
purely gluonic interactions, i.e., all Wilson coefficients equal to
zero apart from $C^S_{g}$. In this case the relative size of two-body
contributions becomes
\begin{multline}
\label{corr_2b_theta}
2 \frac{f_\pi \F_\pi(0)+f_\pi^\theta\F_\pi^\theta(0)}{f_N A} \\
=-2\frac{\mpi}{\mN}\frac{2}{27f_Q^N}\frac{\F_\pi(0)-\F_\pi^\theta(0)}{A}
=-6\%,
\end{multline}
where the large numerical value of $\F_\pi^\theta(0)$ balances the
large coupling of the nucleon to the gluon operator to produce an
effect of similar magnitude as in Eq.~\eqref{corr_2b}.

These examples demonstrate that the hierarchy implied by the nuclear
structure factors themselves can be upset if enhancements or
suppressions in the coefficients, either the nucleon matrix elements
or the Wilson coefficients, are present: in a similar way as
$\F_\pi^\theta$ appears enhanced compared to the scalar two-body
response $\F_\pi$, but is compensated by a large nucleon matrix
element, the impact of the isovector one-body response is suppressed
by small isospin-breaking effects in the nucleon couplings, see
Eq.~\eqref{scalar_ud}.

While the size of two-body corrections can be enhanced by fine-tuning
the one-body coefficients (the one-response-at-a-time strategy put
forward in Sec.~\ref{sec:overview} corresponds to the case where the
cancellation is complete), the special cases in
Eqs.~\eqref{corr_2b}--\eqref{corr_2b_theta} indicate the size of
effects to be expected in regions of parameter space where no such
cancellations of the leading contributions occur.

We stress that in the general case no interrelations such
as Eqs.~\eqref{corr_2b} or~\eqref{corr_2b_theta} between the different form factors exist,
except for those dictated by QCD, e.g.\ $f_2^{V,N}$ contributing both
to $\F_\pm^{\Phi''}$ and to the radius correction to $\F_\pm^M$. Even
in the very special case of scalar interactions considered in Eq.~\eqref{scalar_ud} the dependence on the Wilson
coefficients fully factorizes only if $f_\pi=\tilde f_\pi$ is assumed,
or if the isovector contribution is neglected. In these extreme cases
the limits on the scattering rate immediately translate to a limit on
the single parameter $f_\pi$ and thereby a fixed combination of
$C_u^{SS}$ and $C_d^{SS}$.

In general, direct-detection experiments are sensitive to several
independent combinations of Wilson coefficients, whose determination
therefore requires a correlated analysis of different targets.
Otherwise, simplified strategies such as constraining one response at
a time amount to considering slices through the parameter space of
Wilson coefficients at the hadronic scale.  This information can then
be transferred to a given new-physics model, keeping in mind the
operator running and mixing to be applied when translating the limits
to BSM
scales~\cite{Hill:2014yxa,Hill:2014yka,Crivellin:2014qxa,Crivellin:2014gpa,D'Eramo:2014aba}.
In the interpretation of such limits one also needs to take into
account that not all coefficients are necessarily independent, e.g.,
for the spin-$1/2$ case considered in this paper $8$ parameters map
onto only $7$ ($4$) Wilson coefficients in the Dirac (Majorana) case.

\section{Summary}
\label{sec:summary}

We have presented a strategy for the analysis of general SI WIMP
scattering off nuclei, keeping all terms that lead to a coherent
contribution of nucleons in nuclei and appear up to third order in the
power counting of the WIMP--nucleon interaction according to ChEFT.
Up to dimension $7$ in an effective Lagrangian for WIMP and
Standard-Model fields, scalar and vector interactions on the nucleon
side can give rise to coherent enhancements.  Our analysis shows that
the leading corrections to the standard SI response are the isovector
counterpart and the coherent contribution of WIMPs interacting with
two nucleons (two-body currents).  For a more detailed analysis, the
next corrections to be included are momentum-dependent corrections to
the nucleon form factors as well as the quasi-coherent response associated
with the nucleon spin-orbit operator. The latter only contributes in
the case of vector interactions. Therefore, it could potentially be
used as a tool to experimentally discriminate between the scalar and
vector channels.

Overall, we have found that a generalized SI scattering cross
section including the dominant coherent corrections
depends on $8$ parameters ($4$ in a minimal extension),
which in principle can be fixed by experiments performed with
different nuclear targets, and we have discussed how to constrain these
parameters in direct-detection experiments.  For the case of WIMPs
scattering off xenon isotopes, we have provided parameterizations of
all relevant one- and two-body nuclear responses based on
state-of-the-art nuclear shell-model calculations.  These can be
directly used for a general SI analysis of direct-detection
experiments, considering, e.g., one response function at a time.
In particular, our results show that direct-detection experiments
are sensitive to additional BSM physics than the one coupling
constrained in present standard SI analyses, and thus impose additional 
restrictions on the parameter space of a given new-physics model.

\begin{acknowledgments}

We thank Laura Baudis, Silas R.\ Beane, Vincenzo Cirigliano, Michael
L.\ Graesser, Richard J.\ Hill, Martin J.\ Savage, and Mikhail P.\ Solon for helpful discussions.  
This work was supported in part by the US DOE Grant No.\ DE-FG02-00ER41132,
the ERC Grant No.\ 307986 STRONGINT, the DFG through Grant SFB 1245,
the Max-Planck Society, and JSPS Grant-in-Aid for Scientific Research
No.\ 26$\cdot$04323. JM was supported by an International Research
Fellowship from JSPS. We thank the Institute for Nuclear Theory at the
University of Washington for its hospitality and the US DOE for
partial support during the completion of this work.

\end{acknowledgments}

\end{document}